# Energy Harvesting for Self-Powered Microsystems: A Critical Review of Materials, Power Management, and System Integration

Lyuye Lin[1],Feng Zhifu[1], Ermanno Miele[2], Giuseppe Cantarella[3], Elena Degoli[3], Diego Angeli[3], Lethy Krishnan Jagadamma[4], Feng Gao[5], Michele Magno[6], Ivano Eligio Castelli[7], Tommaso Ongarello[8], Chunbo Liu[9], Roman Krahne[1], Denis Garoli[1,3*] and Remo Proietti Zaccaria[1]

[1] Istituto Italiano di Tecnologia, Via Morego 30, 16163 Genova (Italy)

[2] University of Cambridge, Cavendish Laboratory, Cambridge (UK)

[3] Università degli Studi di Modena e Reggio Emilia, Modena (Italy)

[4] School of Physics and Astronomy, University of St Andrews, North Haugh, St Andrews KY16 9SS (UK)

[5] Department of Physics, Chemistry and Biology (IFM), Linköping University, Sweden

[6] Department of Information Technology and Electrical Engineering, ETH, Zurich (CH)

[7] Technical University of Denmark (DTU) - Department of Energy Conversion and Storage Autonomous Materials Discovery, Agnes Nielsens Vej, 301, 108 2800 Kgs. Lyngby (Denmark)

[8] EssilorLuxottica Smart Eyewear Lab, EssilorLuxottica, 20121 Milano, Italy

[9] Dept. of Mechanical and Electrical Engineering, Henan University of Technology, China (CN)

Corresponding author: Prof. Denis Garoli – denis.garoli@unimore.it

## Abstract

The relentless proliferation of the Internet of Things (IoT), wearable bioelectronics, and cyber-physical infrastructure has rendered the conventional electrochemical battery the single most prohibitive bottleneck to long-term, maintenance-free autonomous microsystems. Energy harvesting, i.e. the conversion of ambient mechanical, thermal, and radiative energy into usable electrical power, has consequently emerged as a transformative paradigm to realize perpetual, battery-independent operation. This review critically synthesizes the most significant advances in energy harvesting technologies over the recent period, with a focus on triboelectric nanogenerators (TENGs), piezoelectric and pyroelectric transducers, indoor photovoltaics, radio-frequency (RF) rectennas, and their multi-source hybrid integrations. We highlight paradigm-shifting breakthroughs, including liquid-solid TENGs that eliminate mechanical wear, nonlinear piezoelectric oscillators that broaden operational bandwidth by over 300%, machine-learning-accelerated material discovery for high charge-density dielectrics, and multiband and broadband RF harvesting enabled by metamaterial architectures. Crucially, we move beyond conventional materials-centric narratives to critically interrogate the "unseen" system-level bottlenecks: ultra-low-voltage cold-start power management integrated circuits (PMICs), the impedance-matching challenges of hybrid energy sources, and the persistent degradation of micro-supercapacitors and thin-film batteries under realistic field conditions. Through detailed case studies in structural health monitoring, wearable cardiac patches, smart glasses, and in-pipe sensor networks, we demonstrate that while harvester efficiencies are approaching theoretical limits, the practical translation to perpetual IoT demands a holistic, co-optimized approach encompassing materials, power electronics, and energy storage. We conclude by identifying the critical gaps, including the lack of standardized testing protocols, the urgent need for 10-year operational reliability, and the promise of AI-driven adaptive power management, and project a roadmap toward the next decade of self-powered microsystems.

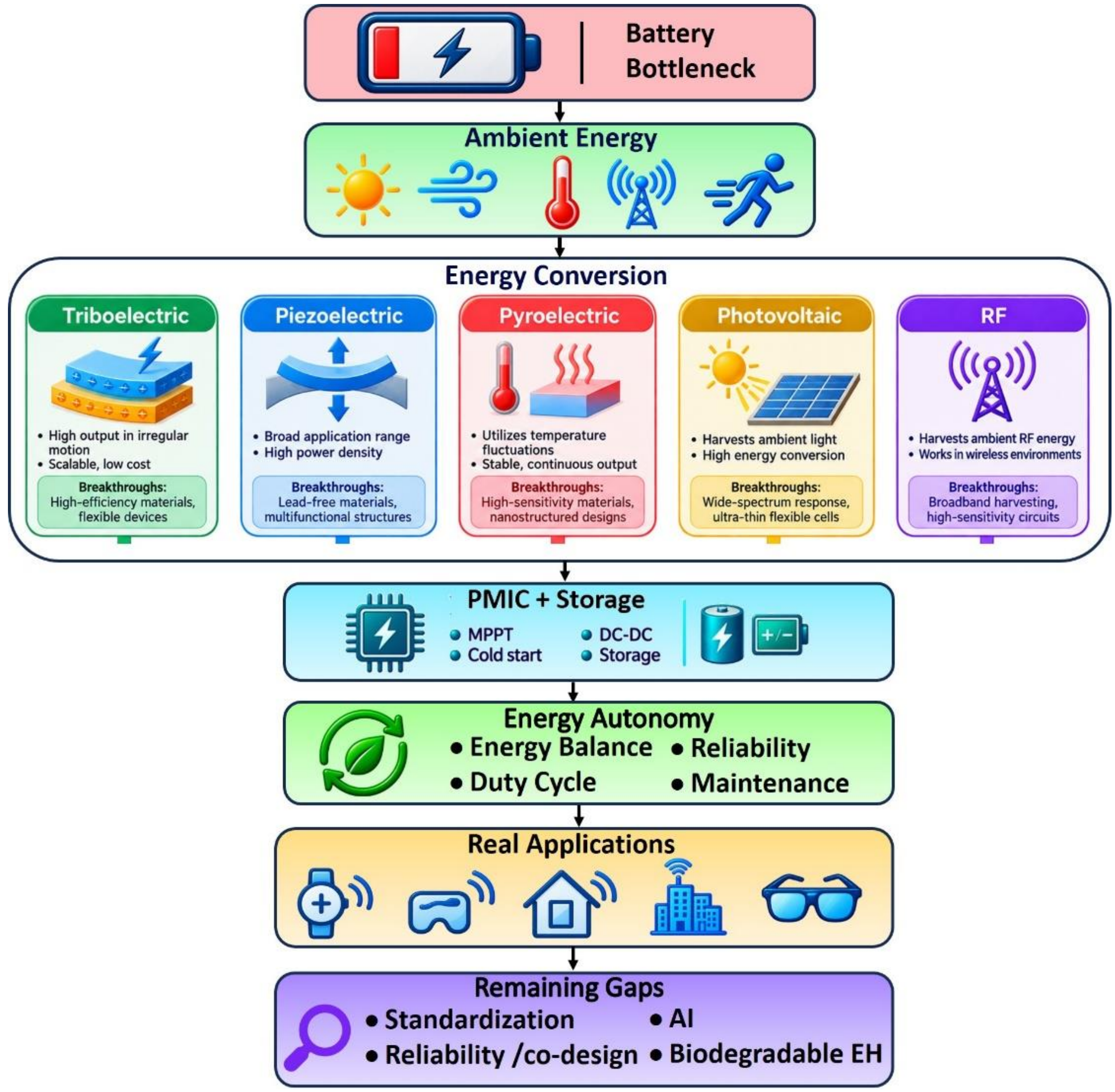

**Figure 1. Graphical Abstract.** Schematic overview of the energy harvesting ecosystem for self-powered microsystems. Five primary harvesting modalities (triboelectric, piezoelectric, pyroelectric, photovoltaic, and RF) are integrated with power management ICs (PMICs) and energy storage elements (supercapacitors and thin-film batteries) to achieve perpetual, battery-independent IoT operation.

## 1. Introduction

The rapid proliferation of the IoT, wearable -monitoring devices, structural health monitoring (SHM) systems, and distributed environmental sensors is driving unprecedented demand for large-scale, long-lived sensor networks. With approximately hundreds of million connected devices envisioned by 2030, the widespread deployment of autonomous micro-sensors for smart cities, Industry 4.0, precision agriculture, and other emerging applications will impose unprecedented power-supply, maintenance burden, and long-term operational sustainability [1–6]. At present, the vast majority of such devices depend on conventional electrochemical batteries. Nevertheless, this reliance presents a fundamental and growing bottleneck to long-term sustainability and autonomy. Batteries feature finite service lifespans, necessitate costly and logistically burdensome maintenance for replacement in remote or hard-to-reach locations, and pose substantial environmental hazards stemming from toxic heavy metals and non-biodegradable packaging material [6–9]. In the context of biomedical implants, battery replacement even requires secondary surgical interventions, introducing significant patient risk [10–12]. Consequently, the pursuit of "perpetual" or "self-sustaining" micro-power sources has emerged as one of the most critical frontiers in microelectronics and materials science. Energy harvesting, i.e. the process of capturing ambient, otherwise wasted energy from the environment and converting it into usable electrical power, offers the most viable pathway to achieving this autonomy [8,13]. The ambient environment is replete with untapped energy reservoirs, broadly categorized into four physical domains: mechanical (vibrations, human motion, wind, acoustic waves, and fluid flow), thermal (waste heat gradients and temperature fluctuations), radiative (ambient light, radio-frequency (RF) electromagnetic waves), and chemical (moisture, enzymatic reactions) [6,14–16]. While early research in this domain focused predominantly on improving the peak power output of individual transducers, the recent trajectory of the field has shifted significantly toward addressing the "real-world" integration challenges [17]. Importantly, superior transducer-level performance does not necessarily translate into system-level energy autonomy. Ambient energy is intrinsically intermittent and considerable losses occur through rectification, impedance matching, power conversion, and energy storage before the harvested energy reaches the load. Therefore, the practical challenge is shifting from maximizing harvested power toward optimizing time-average net usable energy delivered to the microsystem. From this viewpoint, the contemporary research narrative is no longer defined by material-level efficiency alone. Instead, it is shaped by three critical, interdependent pillars: form-factor adaptability (flexibility, stretchability, and biocompatibility) [11,12,18–21], multi-source synergy (hybrid harvesters that capture energy from multiple ambient sources simultaneously) [14,15,22–24], and power-management intelligence (ultra-low-power integrated circuits capable of cold-starting and maximizing energy transfer under highly fluctuating input conditions) [8,25,26]. This review synthesizes most recent advances across these pillars, with a particular emphasis on triboelectric, piezoelectric, pyroelectric, photovoltaic, and radio-frequency (RF) energy harvesting, highlighting emerging strategies that address their respective limitations in bandwidth, durability, adaptability, and system integration. We critically assess the state-of-the-art, identify the persistent impedance-matching and durability bottlenecks that hinder practical deployment, and project the future roadmap toward battery-independent microsystems [27]. The overall source-to-load architecture, spanning ambient energy conversion, power management, energy storage, and self-powered microsystem operation, is summarized in Fig. 1. To provide a unified comparison for the technology-specific discussions that follow, Table 1 compares the principal energy-harvesting technologies in terms of their energy sources, representative electrical outputs, conversion characteristics, form-factor compatibility, and key limitations. The comparison reveals that energy-harvesting technologies occupy distinct operating regimes rather than a single performance hierarchy, with their practical suitability ultimately determined by resource availability, attainable power, electrical impedance, and application-specific constraints. Because efficiency definitions and measurement boundaries differ among harvesting modalities, the peak-efficiency values summarized in Table 1 should be interpreted as technology-specific benchmarks rather than directly comparable system-level efficiencies.

**Table 1. Comprehensive Comparison of Energy Harvesting Modalities**

| Parameter | Triboelectric (TENG) | Piezoelectric (PEH) | Pyroelectric | Indoor Photovoltaic | RF Rectenna | Thermoelectric (TEG) |
|---|---|---|---|---|---|---|
| **Energy Source** | Contact / mechanical motion [28–45] | Strain / vibration [58–73] | Temporal ΔT [76–85] | Indoor light [100,106–110] | Ambient RF [102–105,111,112] | Spatial ΔT [76–79] |
| **Typical Output Power** | μW–mW; >10.8 W $m^{-2}$ peak reported [30,31] | μW–mW; up to 140 mW [59] | nW–μW; 7.4 mW $m^{-2}$ in solar-driven hybrid [85] | Spectrum-dependent; 106.25 μW $cm^{-2}$ predicted [107] | Weak-input-dependent; 562 μW $unit^{-1}$ at 40 μW $cm^{-2}$ [102] | μW–mW; ΔT-dependent [76–79] |
| **Electrical Characteristics** | High V / low I; very high impedance | Moderate–high V; high impedance | High impedance | Low–moderate V; moderate impedance | Low V; rectifier-limited | Low V / higher I; low impedance |
| **Peak Efficiency (2024–2026)** | No universal benchmark; boundary-dependent [30–45] | No universal benchmark; excitation-dependent [58–73] | Up to 27% of Carnot at ΔT = 10 K [80] | 37.4% PCE at 250 lx, 5500 K LED [100] | Up to 97.8% radiation-to-RF at 1.8 GHz [102] | No universal benchmark; ΔT-dependent [76–79] |
| **Typical Frequency Range** | Low-frequency / irregular | Resonant; architecture-dependent | Dynamic thermal cycling | Continuous / intermittent | GHz RF | Steady / slowly varying |
| **Form Factor** | Flexible / stretchable | Rigid / flexible | Material-dependent | Thin-film / flexible | Printed / planar | Predominantly rigid |
| **Cycle Life / Durability** | 10,000 cycles demonstrated for LM-TENG [45] | 30,000 cycles demonstrated for PLLA PENG [72] | Long-term data limited [76–83] | >95% PCE retention after 300 h (DSSC) [101] | Field durability application-dependent [102–105] | Material/application-dependent |
| **Bio-/Environmental Compatibility** | Material-dependent | Biodegradable PLLA/CNC available [72,73] | Material-dependent | Pb concern for PSC; organic alternatives | Material-dependent | Material-dependent |
| **Commercial Readiness (2026)** | Emerging | Established in selected applications | Research-stage | Commercial Si; emerging PSC/DSSC/OPV | Emerging | Established |
| **Key 2024–2026 Breakthrough** | Liquid-solid / LM-TENG; AI-assisted materials [33,39,44,45,49–53] | Nonlinear broadbanding; biodegradable PEHs [64,66–73] | High-field Olsen-cycle conversion [80–83] | >37% indoor PSC; advanced DSSCs [100,101] | Multiband harvesting / passive RF [102–105,111,112] | — |

| **Limitations** | Wear / high impedance | Resonance / brittleness | Requires temporal ΔT | Spectral / illumination variability | Low ambient RF power | Requires sustained ΔT |
|---|---|---|---|---|---|---|

*Note: Reported efficiencies are technology-specific and reflect different conversion boundaries and operating conditions; they should not be interpreted as directly comparable system-level efficiencies.*

Among these modalities, triboelectric nanogenerators provide a particularly instructive example of the opportunities and limitations of ambient mechanical-energy conversion, combining high-voltage output and structural versatility with persistent challenges in charge density, durability, and electrical interfacing.

## 2. Triboelectric Nanogenerators (TENGs): From Material Optimization to System Durability

### 2.1. Fundamentals and the Charge-Density Bottleneck

Since their inception in 2012, TENGs have attracted considerable attention for harvesting low-frequency and irregular mechanical motions (e.g., human joint movement, ocean wave undulation, and finger tapping) into electrical energy because of their broad material compatibility, structural versatility, and simple device architectures [28,29]. The operational mechanism relies on the coupling of contact electrification and electrostatic induction. When two dissimilar materials are brought into physical contact and subsequently undergo relative motion, interfacial charge transfer generates opposite charged surfaces, inducing charge redistribution between the electrodes through an external circuit and thereby generating electrical output. However, despite the remarkable power-generation capabilities demonstrated by TENGs and related triboelectric hybrid nanogenerators under optimized laboratory conditions, with peak power density exceeding 10.8 $W/m^2$ reported for specific device architectures [30,31], the translation of TENGs from academic novelty to industrial viability is persistently obstructed by three interrelated challenges: (i) limitation on surface-charge accumulation associated with air breakdown (Paschen's law), charge leakage and other charge-dissipation mechanisms [32–34]; (ii) the severe mechanical wear and material degradation associated with repetitive solid-solid friction, motivating the development of more mechanically compliant architectures such as liquid-metal-based TENGs [35]; and (iii) the inherent impedance mismatch, as TENGs typically produce high voltage (kV) but ultra-low current (μA), which is incompatible with standard CMOS electronics [36]. Although air breakdown has traditionally been regarded as the primary upper limit mechanism for charge accumulation and is often described using Paschen-type models [32], recent evidence indicates that field emission can also limit charge accumulation before dielectric breakdown, revealing additional physical constraints on the achievable charge density [34]. Recent research landscape has pivoted aggressively to address these bottlenecks through active charge management, durable device architectures, and improved electrical interfacing, yielding several transformative breakthroughs. In this context, Fig. 2. illustrates the fundamental operating mechanism of TENGs and highlights how active charge management can alleviate the charge-density limitations inherent to conventional contact electrification.

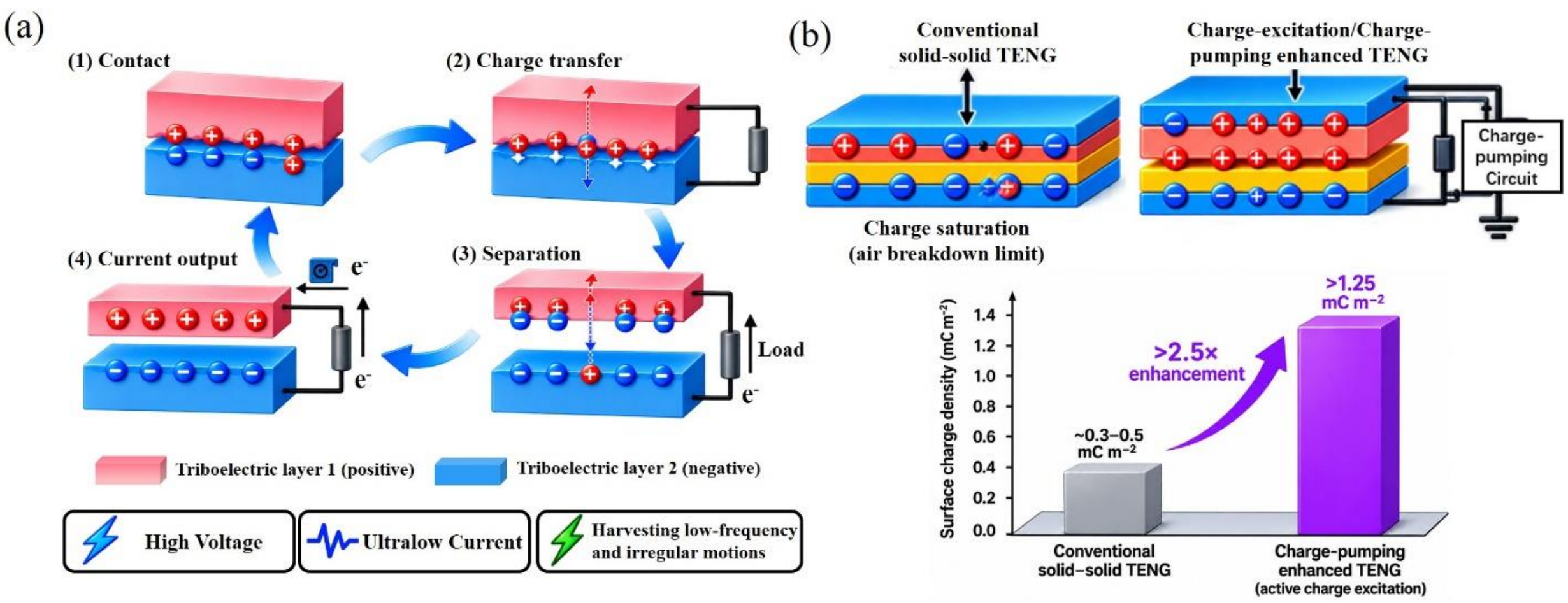

**Figure 2. Triboelectric Nanogenerator (TENG) Operating Principles and Recent Advances.** (a) Fundamental operating principle of a contact–separation TENG, illustrating triboelectric charge transfer upon contact and electrostatic induction during separation [28,29]. (b) Comparison of surface charge density between conventional solid–solid TENGs and charge-pumping-enhanced designs; an effective charge density exceeding 1.25 mC/m² has been demonstrated under ambient conditions [37].

## 2.2. Recent Breakthrough I: Charge-Excitation Strategies and the "Switching" Revolution

To circumvent the inherent limitation of air breakdown and charge saturation, the conventional approach involved encapsulating the TENG in a vacuum or high-pressure inert gas. Although effective in suppressing charge dissipation, this is impractical for distributed IoT and wearable applications. Instead of relying solely on the triboelectric charges generated by friction of the dielectric materials, recent designs incorporate an independent, low-voltage power source to actively inject and accumulate charges on the dielectric surface via synchronized switching mechanism [37,38]. An integrated charge-excitation TENG achieved an effective charge density exceeding 1.25 mC/m² under ambient conditions by combining external/self-charge excitation with voltage multiplication [37]. In another study, Hu et al. showed how the optimization of the electrical matching between a TENG and a voltage-multiplying excitation circuit can increase the transferred charge from 0.04 to 1.54 µC, corresponding to a 38.5-fold enhancement [38]. Subsequently, charge-pumping concepts have been extended to liquid–solid and stacked TENG architectures. An external charge-pumping liquid-solid TENG with charge-storage and extraction capacitors reached a peak power density of 231.8 W/m² with 50uL droplets [39]. For stacked TENGs, a stepwise self-charge pumping strategy increased the short-circuit transferred charge from 50.2 nC to 94.0 nC and uncovered the combined influences of pump voltage, diode reverse-voltage rating and circuit architecture on charge accumulation and electrical breakdown [40]. In parallel, charge-pumping strategies have been extended to multifunctional triboelectric metamaterials, enabling simultaneous enhancement of energy storage and electrical output. A charge-pumping metamaterial increased its charge-storage capacity by 203% and short-circuit charge output by 55.6% through auxiliary-electrode charge accumulation [41]. These advancements demonstrate the evolution of charge excitation from a simple charge-density enhancement strategy toward a broader framework for active charge management involving charge injection, voltage multiplication, charge storage, controlled extraction, and circuit-level modulation. Collectively, these advances mark a transition from passive triboelectric charge generation toward actively managed charge injection, accumulation, storage, and extraction. Recent continuous charge-excitation strategies further demonstrate that dynamically replenishing interfacial charges can overcome the intrinsic surface-charge-density saturation of conventional TENGs, pointing toward a new paradigm of actively regulated triboelectric charge states [33].

### 2.3. Recent Breakthrough II: The Shift to Liquid-Solid Interfaces for Durability

One notable trend in recent TENG research is the increasing shift toward solid-solid to liquid–solid and liquid–liquid triboelectric interfaces. A key advantage of liquid interfaces is the reduction of direct solid–solid frictional heat and associated mechanical wear, potentially improving long-term operational stability of triboelectric devices. At the same time, liquid-based architectures also bring new challenges related to interfacial stability, environmental sensitivity, and system integration.

**Droplet-Based Electricity Generators (DEGs):** Recent DEG research has shifted from simple droplet-impact energy harvesting toward the engineering of liquid-solid interfaces, electrode configurations, and device architectures to improve charge collection, electrical output, output stability and scalability. In particular, surface wettability, droplet spreading dynamics and interfacial adhesion at the interface level have become an important design parameters for regulating charge generation and transfer [42]. In addition, electrode configurations (coplanar and localized electrodes, single- and dual-electrode) and other charge-collection architectures, including patterned wetting surfaces, have also been explored to mitigate parasitic capacitance and improve output performance. Meanwhile, direct-current DEG architectures based on dynamic electric double layers have been developed to bypass conventional rectification, providing a potential route toward more efficient power management and storage [43]. These efforts have progressively extended DEG development from optimizing the electrical response of individual droplets toward scalable device architectures and practical power utilization. For example, recent work demonstrated that tailoring the bottom-electrode area to the droplet spreading area can substantially enhance the average output power of individual DEG cells and enable large-scale 30-cell arrays. By integrating the DEG array with a 400-cell micro-supercapacitor array, the system achieved an energy-storage efficiency of 21.8% and an output power of 81.2 μW, highlighting the importance of electrode optimization, array scaling, and energy-storage integration for practical droplet-based energy harvesting [44].

**Liquid-Metal TENGs (LM-TENGs):** Room-temperature liquid metals, particularly gallium-based alloys such as EGaIn and Galinstan (Ga–In–Sn), are increasingly explored as deformable electrodes in triboelectric energy harvesters. Their high electrical conductivity, fluidity, and deformability allow liquid-metal electrodes to undergo large mechanical deformation and adapt to dynamic triboelectric interfaces while maintaining electrical functionality [35]. Recent device designs have incorporated microstructured liquid-metal electrodes to further improve interfacial contact, output stability, and mechanical durability. Notably, Liu et al. reported on a Galinstan-based TENG incorporating a microarray electrode able to generate an output voltage of ~53 V and to maintain stable performance over 10,000 operating cycles [45], demonstrating the potential of liquid-metal electrodes for mechanically compliant and durable energy harvesting. Representative liquid–solid DEG and liquid-metal TENG architectures illustrating these design concepts are presented in Fig. 3.

It is important to consider that, despite reducing solid–solid friction and mechanical wear, liquid–solid architectures introduce new interfacial, environmental, and chemical reliability challenges. For DEGs, output depends on droplet dynamics, contact conditions, and liquid physicochemical properties; variations in pH and ionic concentration can alter charge transfer and electrostatic screening [46]. Their intermittent, high-voltage output also complicates rectification, power management, and energy storage at scale [44]. For LM-TENGs, surface oxidation can alter interfacial properties, surface tension, and adhesion [35,47], despite demonstrated cycling stability [45]. Thus, the reliability bottleneck shifts from mechanical wear toward interfacial stability, liquid chemistry, environmental robustness, and effective energy utilization.

## 2.4. Recent Breakthrough III: AI-Assisted Triboelectric Material Discovery

Triboelectric material selection has evolved from empirical screening toward quantitative and data-driven design. Standardized triboelectric characterization provides a basis for systematic material comparison [48]. For example. a graph neural network (GNN)-based machine learning (ML) framework achieved 98% classification accuracy, increasing PTFE energy density by 65.7% with Al doping and 85.7% with F doping; a PTFE/Cu system with 7% Ag doping reached 1.12 J/cm² [49]. Interpretable ML further enables rapid prediction and identification of key structural parameters through surrogate modeling and TreeSHAP analysis, reducing reliance on repetitive simulations and experiments [50]. Physics-informed ML incorporates atomic and surface descriptors to improve the accuracy and generalizability of TENG performance prediction [51]. Complementarily, Density Functional Theory (DFT) provides microscopic insights into interfacial charge transfer and electronic structure, supporting physics-based material and interface optimization [52,53]. Collectively, these advances are driving TENG development from empirical screening toward rational, data-driven, and physics-informed design. As illustrated in Fig. 4, GNN-assisted material screening exemplifies this emerging paradigm by linking material descriptors with performance prediction to accelerate the identification of promising triboelectric material combinations.

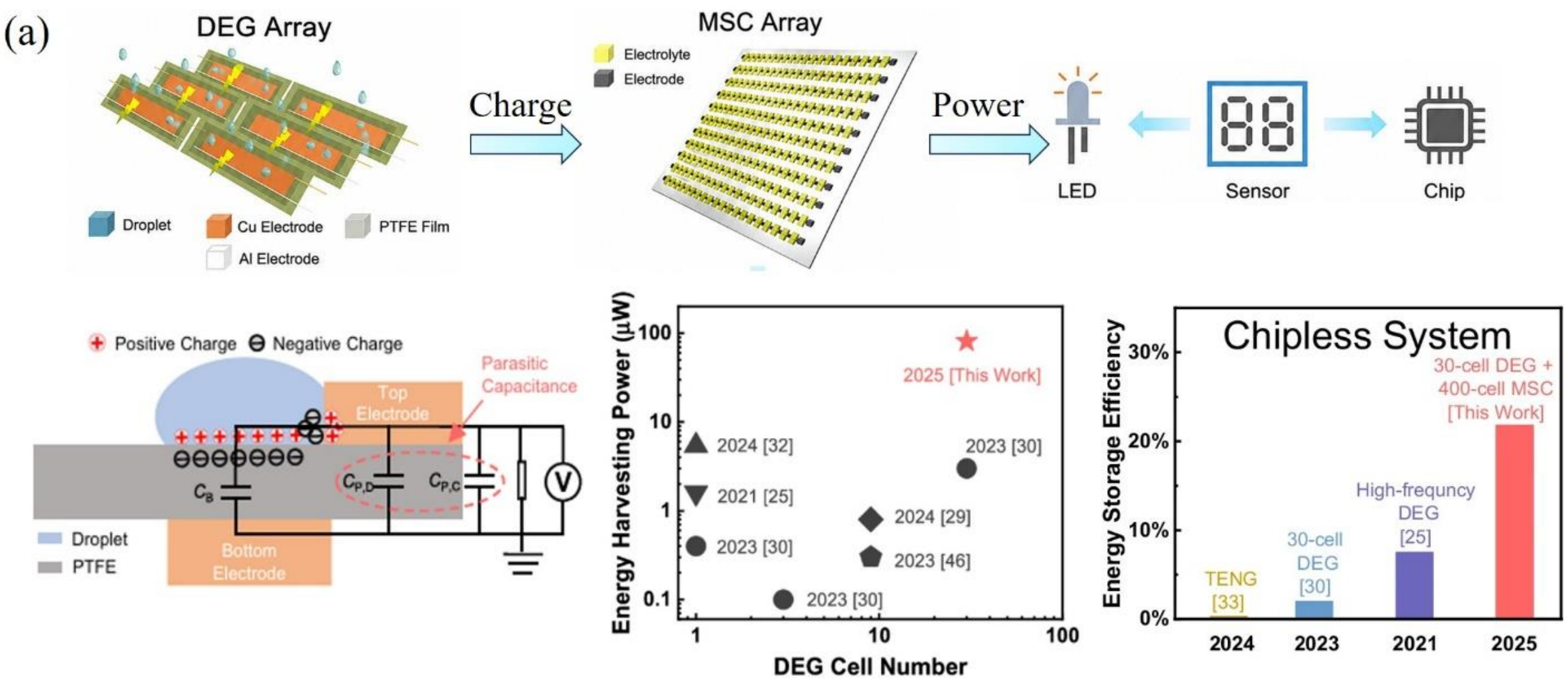

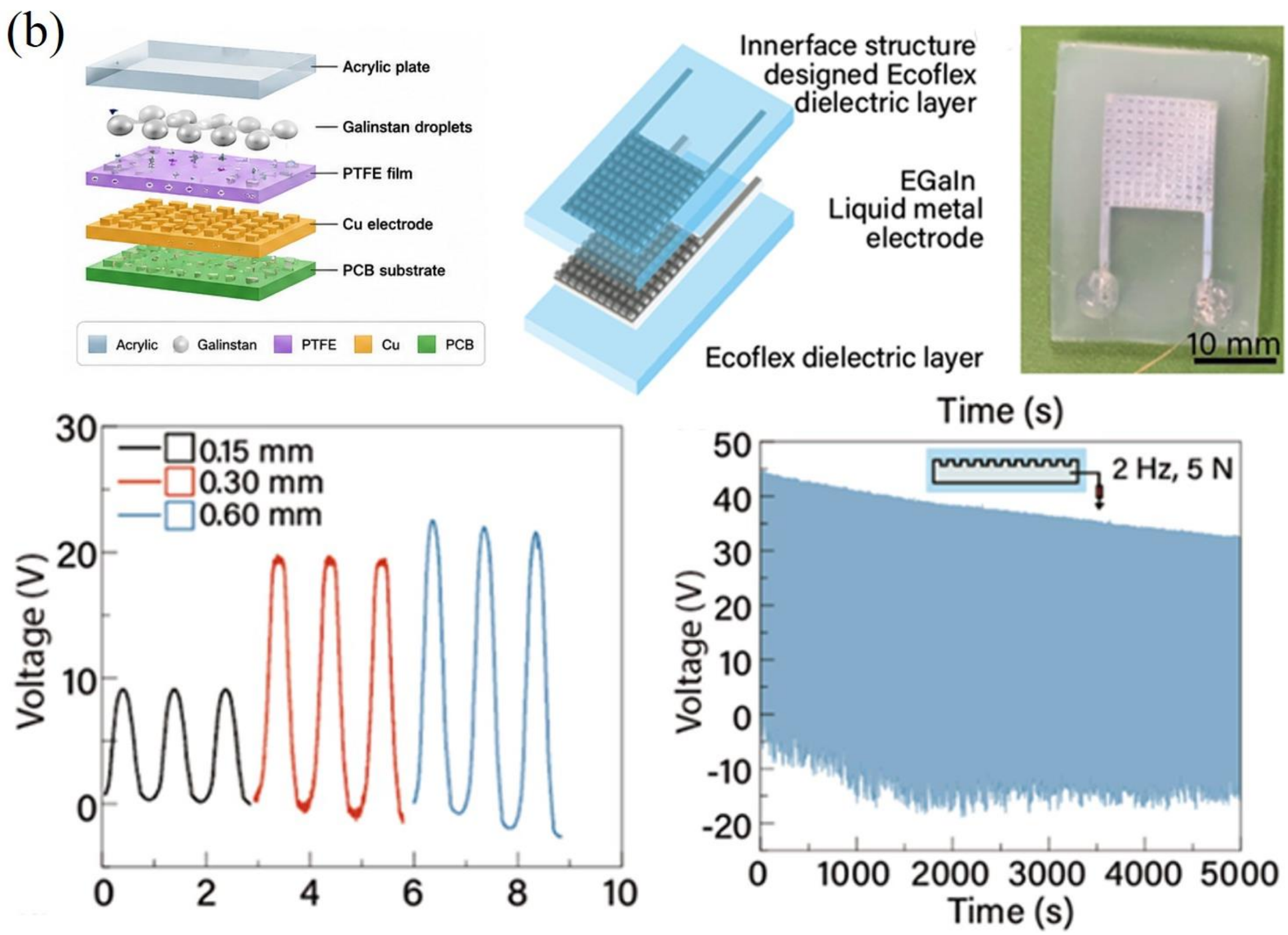

**Figure 3. Triboelectric Nanogenerator (TENG) Operating Principles and Recent Advances.** (a) Liquid–solid droplet-based electricity generator (DEG), in which dynamic liquid–solid interfacial interactions enable scalable energy harvesting; integration with a 400-cell micro-supercapacitor array yielded an output power of 81.2 μW and an energy-storage efficiency of 21.8% [44]. (b) Liquid-metal TENG based on a microstructured Galinstan electrode, demonstrating mechanically compliant operation with an output voltage of approximately 53 V and stable performance over 10,000 cycles [45].

## 2.5. Hybridization: TENGs as Self-Powered Sensors and Active Voltage Triggers

A notable shift in recent TENG research is the transition from using TENGs primarily as standalone power sources toward exploiting their intrinsic electrical characteristics for self-powered sensing and active triggering of other energy-conversion processes [54–56]. Owing to their direct conversion of mechanical stimuli into electrical signals, TENGs can function as self-powered sensors without external bias, enabling force, motion, vibration, and environmental monitoring [54]. For example, a human-body-electrode-enabled DC-TENG generated up to 700 V and 23 μA while enabling wireless human-motion and environmental monitoring [54]. Beyond sensing, the high-voltage and transient output of TENGs can provide active electrical bias or electrostatic-field modulation for complementary energy-conversion processes rather than serving solely as a continuous power source [55,56]. Earlier hybrid TENG–piezoelectric–pyroelectric architectures demonstrated the integration of multiple energy-conversion mechanisms within a single device, achieving a 26.2% enhancement in energy-harvesting efficiency [57]. More recently, Guo et al. reported on triboelectric–photovoltaic field coupling able to increase photovoltaic efficiency from 18.4% to 20.84% [55], while Hu et al. reported on triboelectric–photocatalytic system employing the external bias generated by a wind-driven TENG to reinforce the interfacial electric field of a photocatalyst, achieving a hydrogen-evolution rate of 935.1 μL $m^{-1}$ $m^{-2}$, 34.5 times that of conventional systems [56]. Collectively, these advances highlight a functional shift from TENGs as primary power sources toward active transducers that enable self-powered sensing and trigger or enhance complementary energy-conversion processes.

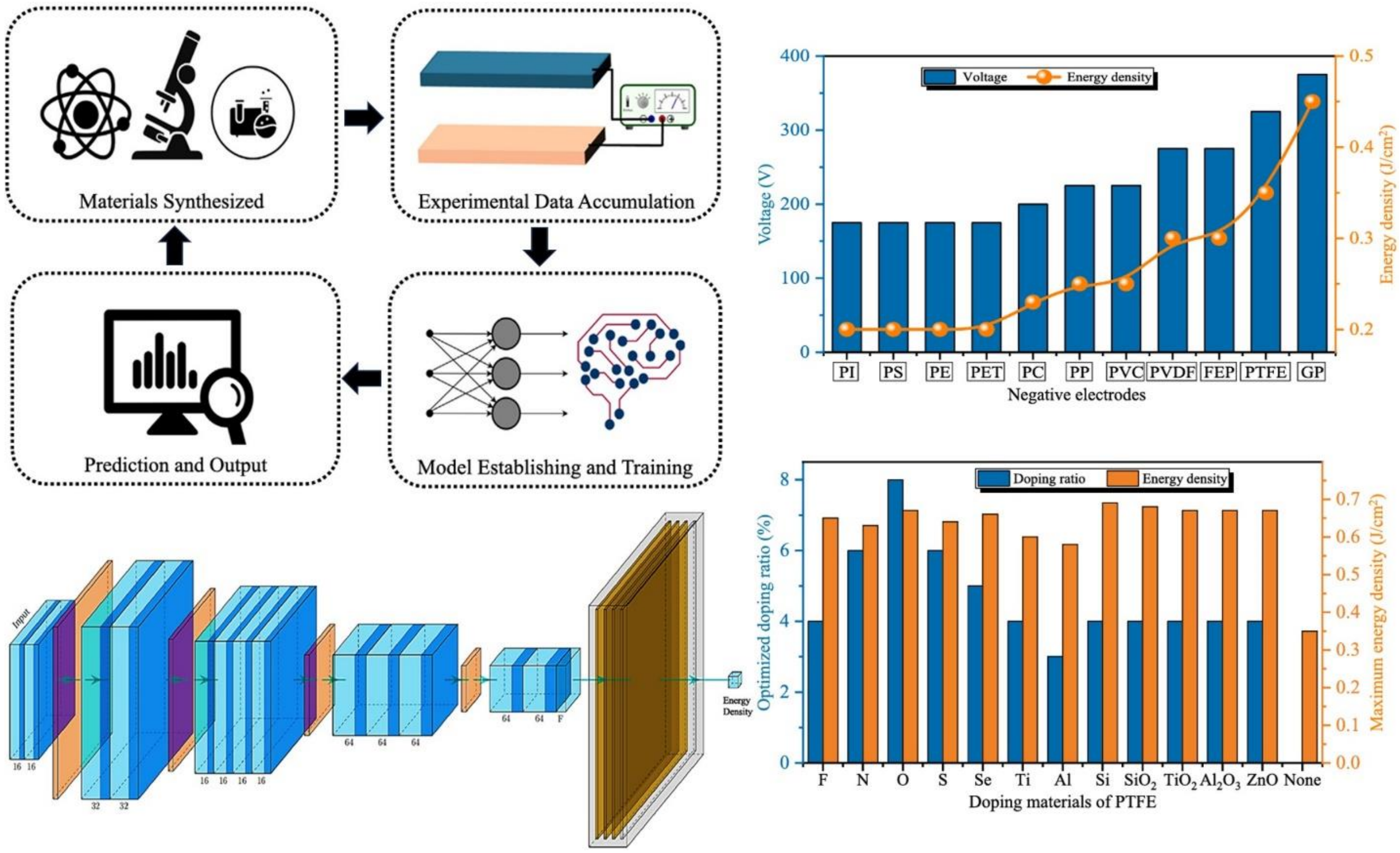

**Figure 4. Triboelectric Nanogenerator (TENG) Operating Principles and Recent Advances**. Machine-learning-assisted triboelectric material discovery using a graph neural network (GNN)-based framework, achieving 98% accuracy in output-performance prediction and identifying substantial energy-density enhancements of 65.7% and 85.7% for Al- and F-doped PTFE, respectively [49].

## 3. Piezoelectric Energy Harvesting: Overcoming the Bandwidth and Brittleness Barriers

### 3.1. The Persistent Spectral Mismatch

PEHs, predominantly based on Lead Zirconate Titanate (PZT), Aluminum Nitride (AlN), and Polyvinylidene Fluoride (PVDF), remain the gold standard for converting ambient mechanical strain into electrical charge [58]. Their operational principle, the direct piezoelectric effect, offers high electromechanical coupling coefficients ($k^2$) and instantaneous power densities that can reach the mW range, sufficient to directly power a duty-cycled wireless sensor node [59–61]. For example, a strongly coupled piezoelectric stack harvester delivered 140 mW at 0.5 g and 157 Hz, with a 10 mW operating bandwidth of 24 Hz, while a railway-oriented harvester achieved 24.5 mW under measured track vibrations [59,60]. However, the Achilles' heel of conventional PEHs is their inherent linear resonant behavior [62,63]. A classic cantilever-based PEH exhibits a sharp frequency response, with maximum power generated when the ambient vibration frequency approaches its natural resonance frequency (fn) [58,62,63]. In real-world environments, such as human motion, infrastructure, and rotating machinery, vibrations are often stochastic, broadband, and time-varying. Even modest frequency detuning can substantially reduce the harvested power [62,63]. Consequently, a major research thrust in recent years has been the systematic effort to "flatten" and "broaden" this resonance peak through multimodal, nonlinear, coupled-resonance, and frequency-tuning strategies [62,64,65]. For instance, Wang et al. [64] developed a spring-based bistable energy harvester (SBEH), in which the nonlinear restoring force generated by pre-compressed springs creates a bistable potential landscape and enables large-amplitude inter-well oscillations over a broader frequency range, as illustrated in Fig. 5. Two configurations were introduced to accommodate different excitation regimes. Under suprathreshold excitation, the SBEH-sup configuration substantially broadens the response compared with the conventional linear energy harvester (CLEH). More importantly, under subthreshold excitation, the SBEH-sub configuration incorporates magnetic regulation to dynamically modify the potential-energy barrier, thereby reducing the threshold for inter-well transitions and sustaining large-amplitude oscillations at lower excitation levels. This potential-well modulation broadens the operating bandwidth from 0.8 to 6.7 Hz (~8.4-fold) and increases the maximum power output by approximately 269% (~2.69-fold) relative to the SBEH-sup configuration. These results demonstrate that dynamically regulating the potential barrier, rather than relying solely on a fixed bistable potential landscape, can substantially improve broadband energy harvesting under weak environmental vibrations. [64] More recently, a multi-frequency nonlinear PEH achieved approximately 35 mW over 25–40 Hz, demonstrating the potential of broadband designs to simultaneously maintain high output power and extend the usable frequency range [65].

### 3.2. Nonlinearity via Magnetic Bistability and Frequency Up-Conversion

A major strategy for overcoming the narrowband response of conventional PEHs is to deliberately introduce nonlinear dynamics into the mechanical architecture. Magnetic interactions provide a tunable means of generating nonlinear restoring forces and bistable or multistable potential landscapes, thereby promoting large-amplitude inter-well motion and broadening the effective harvesting bandwidth [66–68]. A complementary strategy is frequency up-conversion (FUC), in which low-frequency environmental excitation is converted into higher-frequency oscillations of the piezoelectric element through mechanical or magnetic plucking [69–71]. Accordingly, recent advances can be broadly categorized into two complementary routes: magnetic bistability for broadband resonance enhancement and FUC for harvesting ultra-low-frequency excitation.

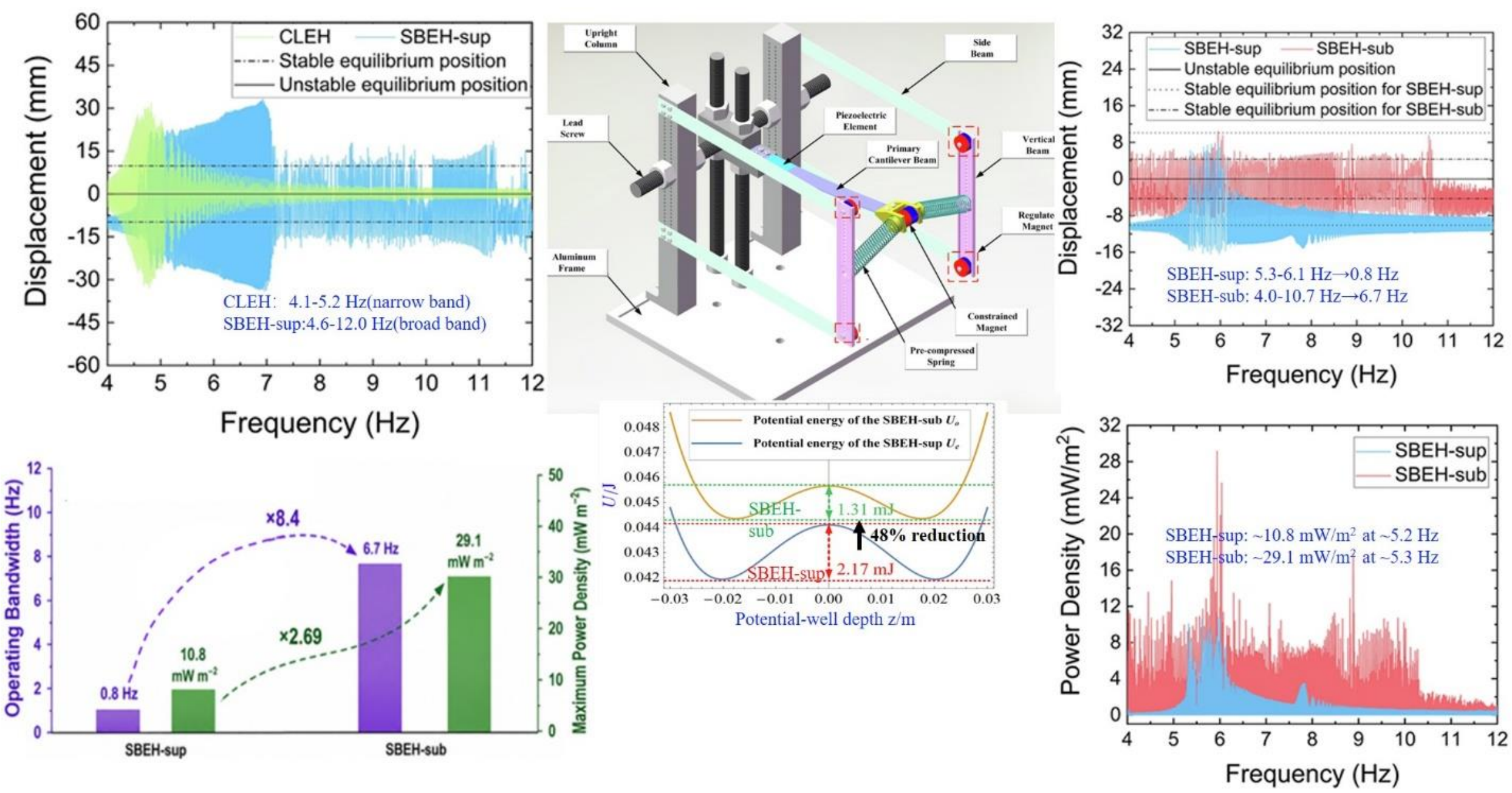

**Figure 5. Spring-based bistable piezoelectric energy harvesting**. Magnetic regulation lowers the bistable potential barrier, enabling inter-well oscillations under subthreshold excitation and enhancing the operating bandwidth (~8.4-fold) and maximum output power (~2.69-fold). Adapted with permission from Ref. [64].

**Magnetic Bi-stability:** The effectiveness of magnetic nonlinear stiffness has been demonstrated across a range of bistable and multistable PEH architectures. As presented in Fig. 6, a complementary strategy based on magnetic nonlinearity was demonstrated by Wang et al. [66]. Their multi-magnet-coupled bistable piezoelectric energy harvester (MC-BPEH) employs multiple adjustable magnets to reshape the magnetic field and strengthen the nonlinear negative-stiffness effect, thereby creating tunable bistable potential wells. Under 0.5g excitation, the four-magnet configuration achieved a maximum peak-to-peak voltage of 114 V, an effective bandwidth of 4.5 Hz, and a maximum output power of 12 mW, corresponding to approximately 30%, 67%, and 78% improvements, respectively, compared with the two-adjustable-magnet configurations[66]. Similar principles were subsequently extended to a stacked tri-stable architecture based on magnetically coupled L-shaped beams, which achieved 1.49 mW at 0.4 g and demonstrated broadband harvesting under low-frequency excitation [67]. Further enhancement was achieved by integrating magnetic bistability with piezoelectric–electromagnetic transduction. Wang et al. reported on low-frequency piezoelectric–electromagnetic harvester where magnetic repulsion was introduced into an M-shaped structure to facilitate inter-well transitions at reduced excitation levels. As a result, the harvesting bandwidth increased by 35.71%, while the piezoelectric and electromagnetic power outputs were enhanced by 57.57% and 55.45%, respectively [68]. Collectively, these studies demonstrate that magnetic bistability can transform the sharp resonance of linear PEHs into a more adaptable nonlinear response, particularly under low-frequency and broadband excitation.

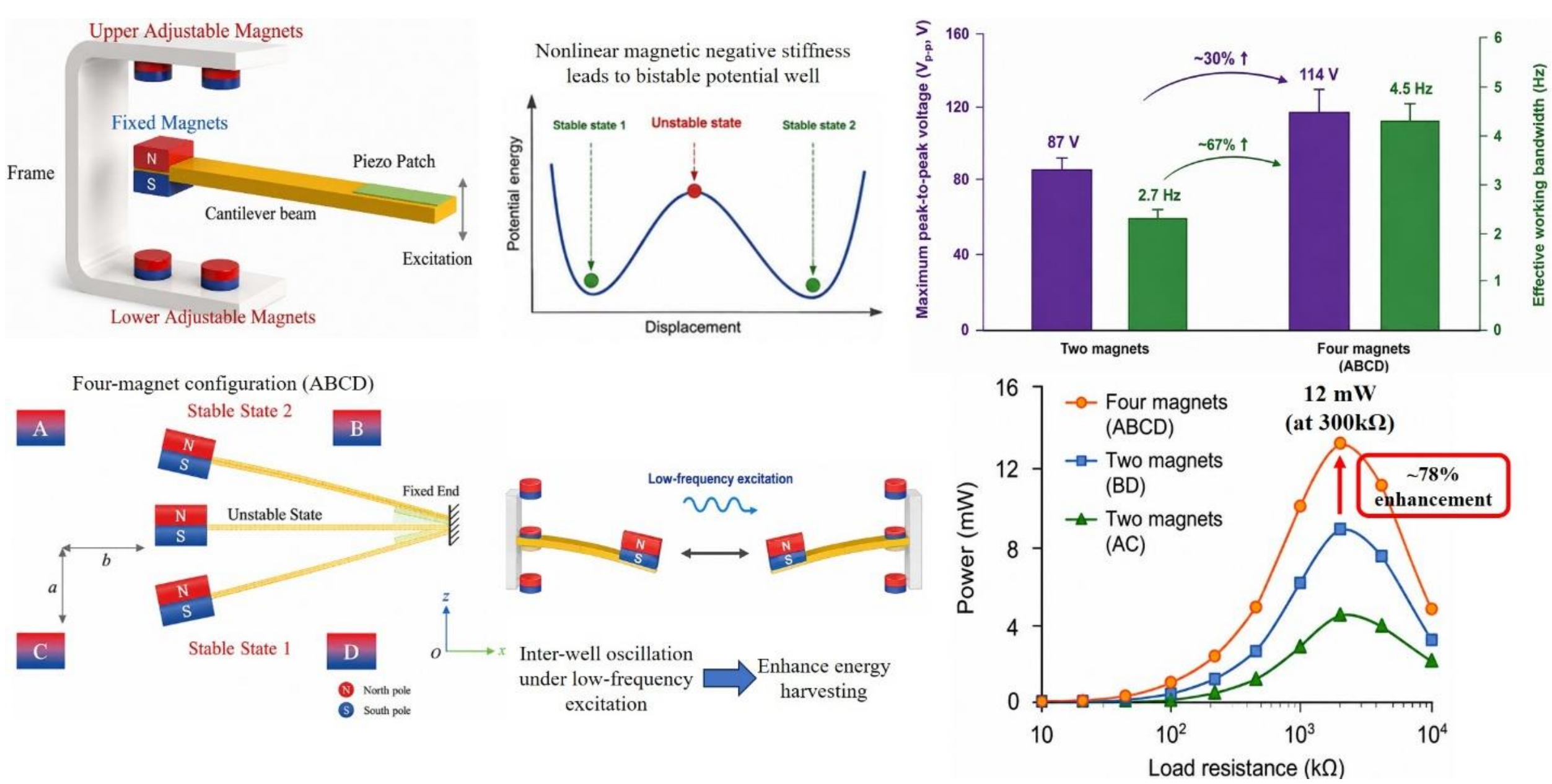

**Figure 6. Multi-magnet coupled bistable for enhanced piezoelectric energy harvesting.** Multiple adjustable magnets introduce nonlinear magnetic negative stiffness and create bistable potential wells, facilitating large-amplitude inter-well oscillations under low-frequency excitation. The four-magnet configuration achieves a maximum peak-to-peak voltage of 114 V and an effective working bandwidth of 4.5 Hz at 0.5g, representing increases of approximately 30% and 67%, respectively, over the two-magnet configuration. Adapted with permission from Ref. [66].

**Frequency Up-Conversion (FUC):** For ultra-low-frequency excitation, a key limitation of conventional PEHs is the mismatch between the excitation frequency and the harvester's intrinsic resonance. Frequency up-conversion addresses this constraint by using a low-frequency mechanical input to trigger high-frequency oscillations of the piezoelectric element. In magnetic-plucking configurations, a slowly moving magnet periodically interacts with a magnet coupled to the piezoelectric beam, inducing rapid beam oscillations near its resonant modes. The effectiveness of FUC has been demonstrated under low-frequency excitation. Shen et al. showed how magnetic-plucked bistable PEH incorporating a nonlinear energy sink exhibited inter-well periodic and chaotic response regimes; at 0.25 g excitation, an average power of 4.71 mW was obtained over 3.30–4.31 Hz during forward frequency sweeping [69]. From a mechanistic perspective, magnetic-plucking experiments have shown that increasing the interaction velocity progressively activates the first bending mode of the piezoelectric bimorph, providing direct evidence of frequency conversion from slow external motion to rapid structural oscillations [70]. Beyond magnetic plucking, impact-induced FUC has been integrated with hybrid piezoelectric–electromagnetic energy harvesting. As shown in Fig. 7, Su et al. [71] further developed an eccentric-rotor-based hybrid harvester in which low-frequency excitation drives rotational motion and electromagnetic generation, while periodic rotor–beam impacts excite high-frequency free oscillations of a fan-folded piezoelectric beam. At 4 Hz, the piezoelectric and electromagnetic components delivered maximum powers of 2.8 and 26.4 mW, respectively. These studies demonstrate that FUC can decouple low-frequency environmental motion from the higher-frequency oscillations required for efficient piezoelectric transduction, providing a complementary strategy to bistability for harvesting low-frequency and irregular mechanical energy.

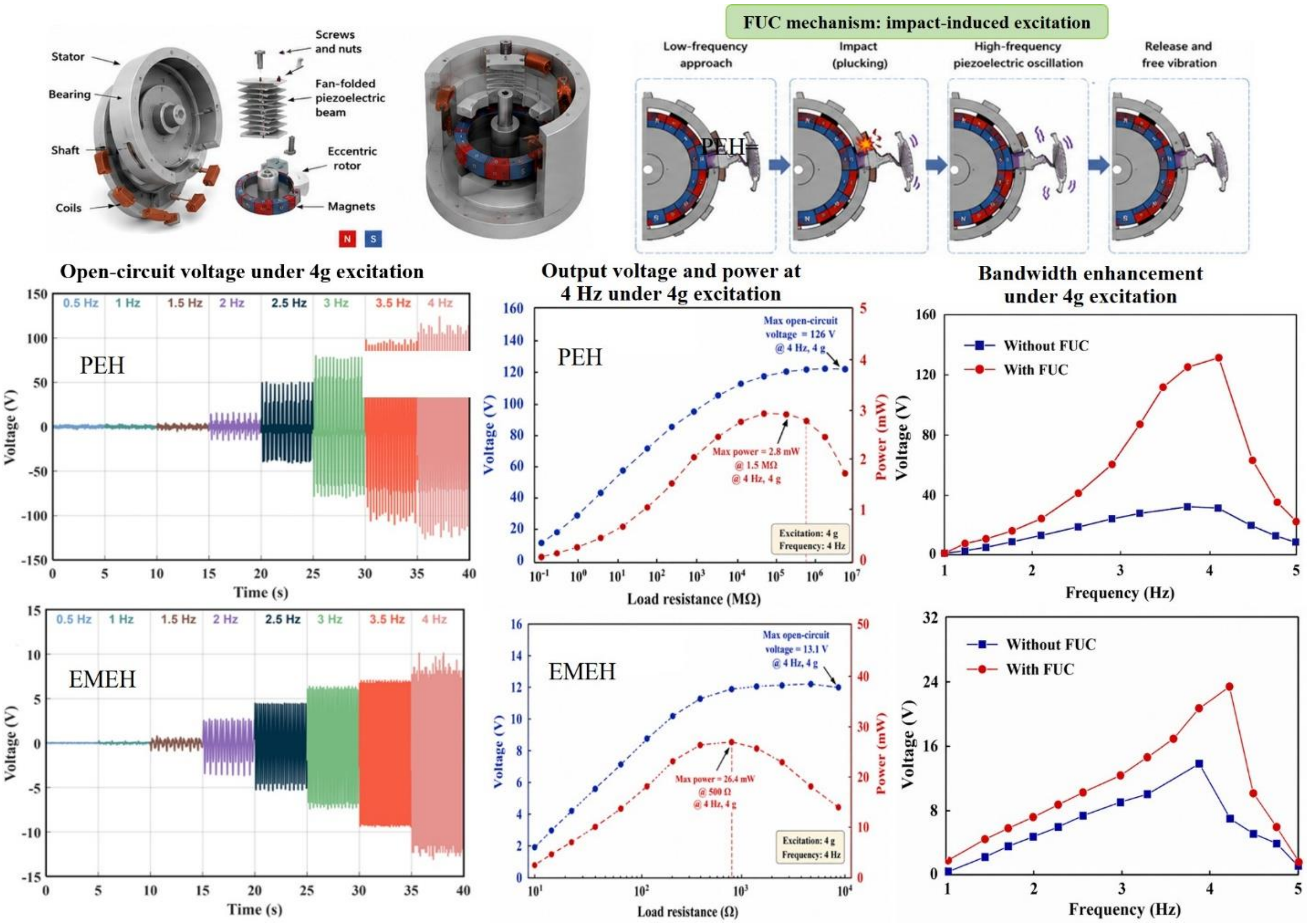

**Figure 7. Impact-induced frequency-up conversion in a hybrid piezoelectric–electromagnetic harvester.** Rotor–beam impacts convert low-frequency motion into high-frequency piezoelectric oscillations, broadening the operating response and yielding maximum PEH and EMEH powers of 2.8 and 26.4 mW, respectively, at 4 Hz. Adapted with permission from Ref. [71].

### 3.3. The Rise of Biodegradable and Flexible Piezoelectrics (PLLA and Cellulose)

Despite the excellent piezoelectric performance of PZT, its rigidity, fragility, and toxic lead content have limited its applicability in wearable, bio-integrated, and implantable energy harvesters. To address these limitations, increasing attention has been directed toward flexible, biodegradable, and biocompatible piezoelectric polymers, bio-derived materials. Among these, PLLA-based polymers and cellulose-derived piezoelectrics have emerged as promising material platforms for next-generation sustainable and transient PEHs.

**Poly-L-lactic Acid (PLLA):** PLLA has emerged as an attractive material for transient and biomedical PEHs owing to its biodegradability, biocompatibility and mechanical conformability. However, its intrinsically modest piezoelectric response, which is strongly governed by molecular organization and dipole orientation, remains a key limitation. Recent efforts have therefore focused on regulating the molecular and microstructural organization of PLLA to enhance electromechanical conversion. In particular, incorporating cellulose into electrospun PLLA nanofibres regulates crystallinity and increases the amorphous fraction, facilitating the electric-field-induced orientation of C=O dipoles and increasing the longitudinal piezoelectric coefficient ($d_{33}$) from 39.8 to 64.2 pm/V at an optimal cellulose content of 1.5 wt% (Fig. 8) [72]. The resulting biodegradable PENG delivered a maximum open-circuit voltage of 10.3 V and short-circuit current of 261.8 nA, while maintaining stable output over 30,000 loading cycles. Importantly, the cellulose/PLLA film exhibited >93.6% mass loss after 120 days of soil burial, demonstrating that enhanced electromechanical performance can be achieved without sacrificing end-of-life degradability [72]. These existing literatures reveal how molecular and microstructural regulation can mitigate the intrinsic piezoelectric limitations of PLLA while preserving its sustainability, providing a pathway toward high-performance flexible and transient PEHs.

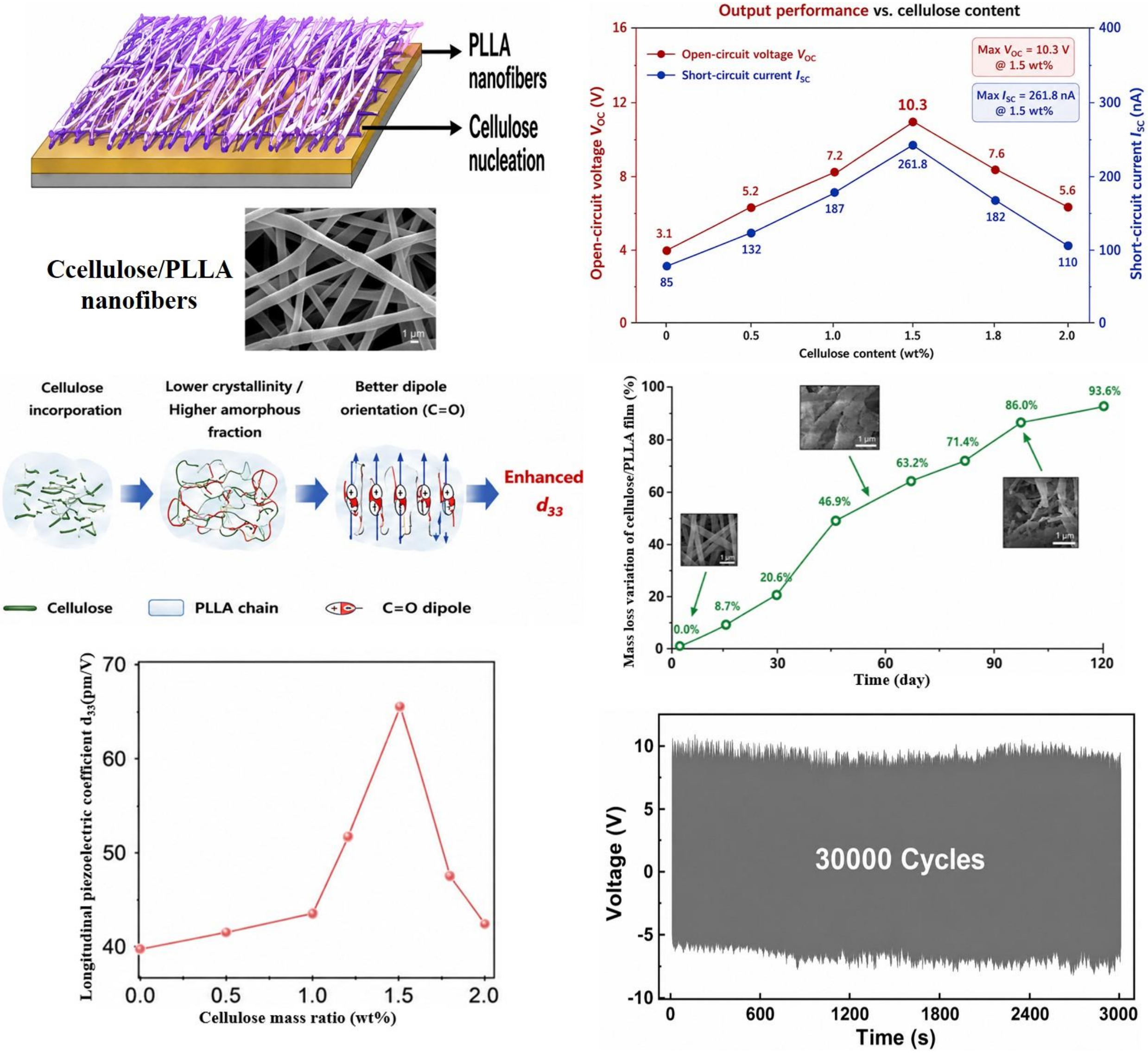

**Figure 8. Cellulose-assisted biodegradable PLLA piezoelectric.** Cellulose incorporation regulates PLLA crystallinity and facilitates dipole orientation, increasing the longitudinal piezoelectric coefficient from 39.8 to 64.2 pm/V. The optimized cellulose/PLLA PENG achieves a maximum open-circuit voltage of 10.3 V, while retaining stable output over 30,000 cycles and exhibiting >93.6% mass loss after 120 days of soil burial. Adapted with permission from Ref. [72].

**Cellulose-Based Piezoelectric:** Beyond PLLA-based systems, cellulose nanocrystals (CNCs) offer a sustainable platform for flexible and biodegradable piezoelectric devices. Recent advances have shifted the emphasis from maximizing the intrinsic piezoelectric response of cellulose toward hierarchical structuring and device-level integration. Ghosh et al. used submicrometre patterning integrated with multilayer stacking to construct fully biodegradable CNC-based piezoelectric arrays, with increasing numbers of active layers progressively enhancing both voltage and current outputs, as described in Fig. 9 [73]. The three-layer architecture achieved a maximum instantaneous power density of 0.6 μW/cm², illustrating how structural integration can translate nanoscale piezoelectricity into enhanced device-level performance. The same platform further enabled gentle-touch pressure sensing (4.2 V/kPa) and exhibited compatibility with cell cultures and implantation, including stable piezoelectric output from devices placed on the epicardium of swine hearts [73]. Collectively, these developments mark a transition from material-level optimization toward hierarchically engineered, application-specific and biodegradable PEHs.

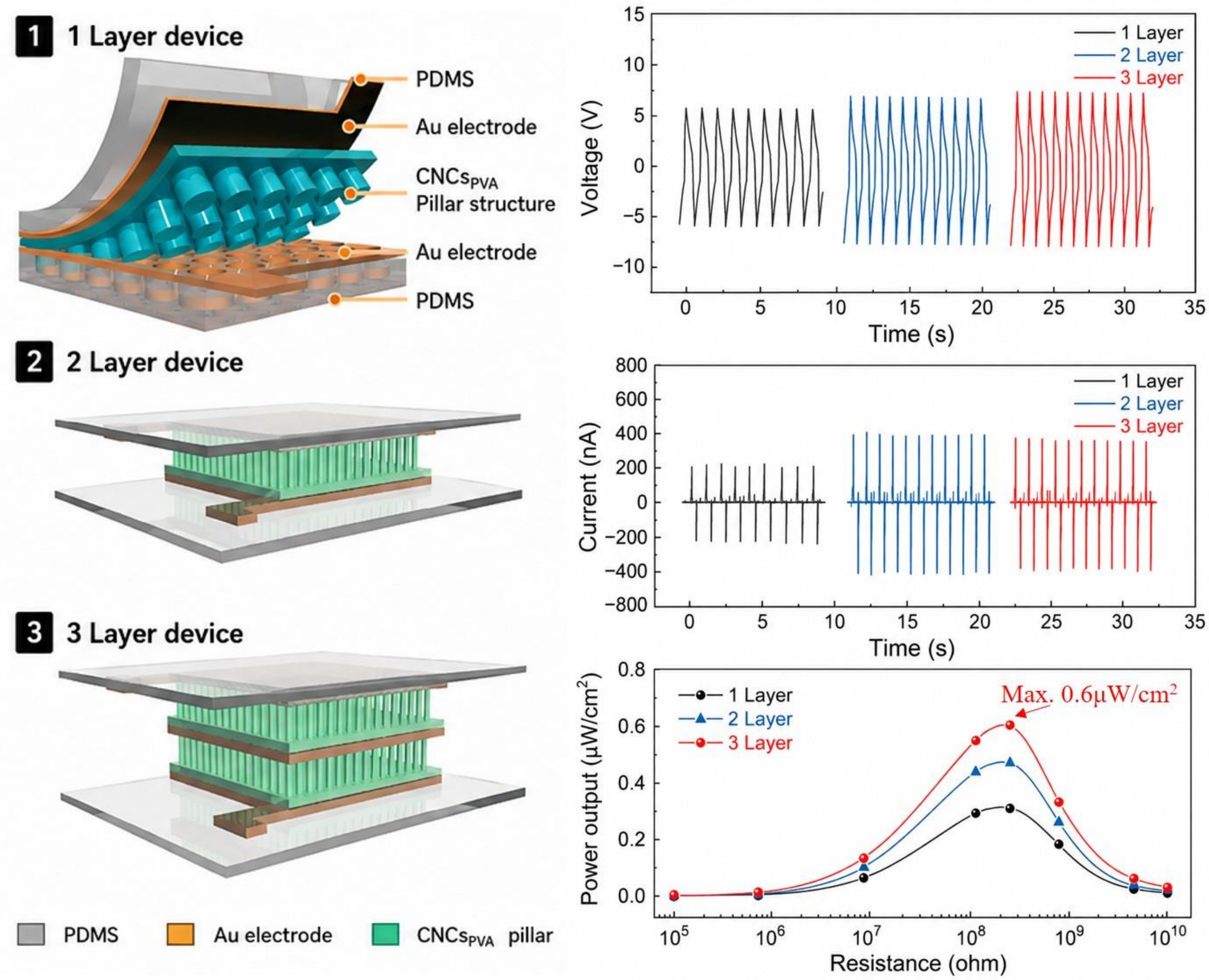

**Figure 9. Hierarchically structured biodegradable nanocellulose piezo-arrays**. Multilayer stacking progressively enhances the piezoelectric voltage and current outputs, with the three-layer architecture achieving a maximum instantaneous power density of 0.6 μW/cm², illustrating the transition from material-level optimization toward integrated biodegradable PEHs. Adapted with permission from Ref. [73].

Importantly, the primary value of biodegradable PEHs may lie less in replacing high-performance PZT for continuous power generation than in enabling applications where biodegradability, mechanical compliance and biocompatibility outweigh maximum power density. Recent implantable piezoelectric systems further illustrated this application-oriented shift, with an i-PENG system demonstrated on a pig heart charging a 100 μF capacitor to 4 V within 13 min for pacemaker operation [74]. Meanwhile, sulfated CNCs have emerged as a promising bio-derived piezoelectric material for energy harvesting and health-monitoring applications [75]. Collectively, these developments point toward an application-driven paradigm in which biodegradable PEHs are targeted primarily at transient electronics, self-powered biomedical sensing and implantable systems.

## 4. Pyroelectric Harvesting: Exploiting Temporal Thermal Gradients

Pyroelectric energy harvesting provides a complementary approach to convert low-grade thermal fluctuations into electricity, obviating the need for a steady spatial temperature gradient [76–78]. In contrast to thermoelectric conversion, operating using a temperature bias across the material, pyroelectric transduction exploits temperature-driven modulations of the intrinsic polarization [76,78,79]. However, the relatively slow and intermittent nature of ambient thermal fluctuations limits the achievable current and energy density [76,77]. This limitation has motivated recent efforts to enhance polarization changes and accelerate thermal cycling through advanced materials, thermodynamic cycles, and thermal-switching architectures [77,79–83].

### 4.1. Differentiating Pyroelectrics from Thermoelectric

Pyroelectric harvesters exploit a fundamentally different thermal driving mechanism from conventional thermoelectric generators (TEGs). As illustrated as Fig. 10(a), whereas TEGs generate a voltage through the Seebeck effect in response to a spatial temperature difference ($\Delta T$), pyroelectric devices respond to a temporal temperature variation ($dT/dt$): a change in temperature alters the spontaneous polarization of the pyroelectric material, inducing compensating charge flow through the external circuit [76–79]. The resulting short-circuit current can be expressed as $I_{\mathrm{sc}} = pA(dT/dt)$, where $p$ is the pyroelectric coefficient and A is the electrode area [76–78]. Consequently, unlike TEGs, which can sustain an output under a stationary temperature gradient, pyroelectric harvesters require time-dependent heating and cooling and are therefore particularly suited to fluctuating thermal sources, including waste-heat fluctuations, body-temperature variations, and periodically heated surfaces. However, direct pyroelectric conversion is intrinsically constrained by the magnitude and rate of the available temperature variation. Under modest thermal fluctuations, the resulting electrical output can therefore remain small; for example, a pyroelectric self-charging cell generated only 0.65 mV after four 20–30–20 °C thermal cycles [84]. Increasing the thermal modulation rate can substantially enhance the electrical response: a solar-driven pyroelectric hybrid device operating at temperature-change rates of approximately 0.5 °C/s achieved a maximum open-circuit voltage of 80 V, a short-circuit current of 0.4 μA, and a maximum power density of 7.4 mW/m² [85]. Beyond simply accelerating thermal cycling, thermodynamic-cycle engineering provides another route to increase the extractable electrical work. As depicted in Fig. 10b, the Olsen cycle couples temperature cycling between $T_{\mathrm{L}}$and $T_{\mathrm{H}}$with electric-field-controlled polarization and depolarization, forming a closed loop in polarization–electric-field (P–E) space whose enclosed area represents the net electrical work generated per cycle [76,77]. These considerations have motivated the development of high-performance pyroelectric materials and actively or passively controlled thermal-switching architectures that maximize polarization changes while accelerating heat exchange [76–79].

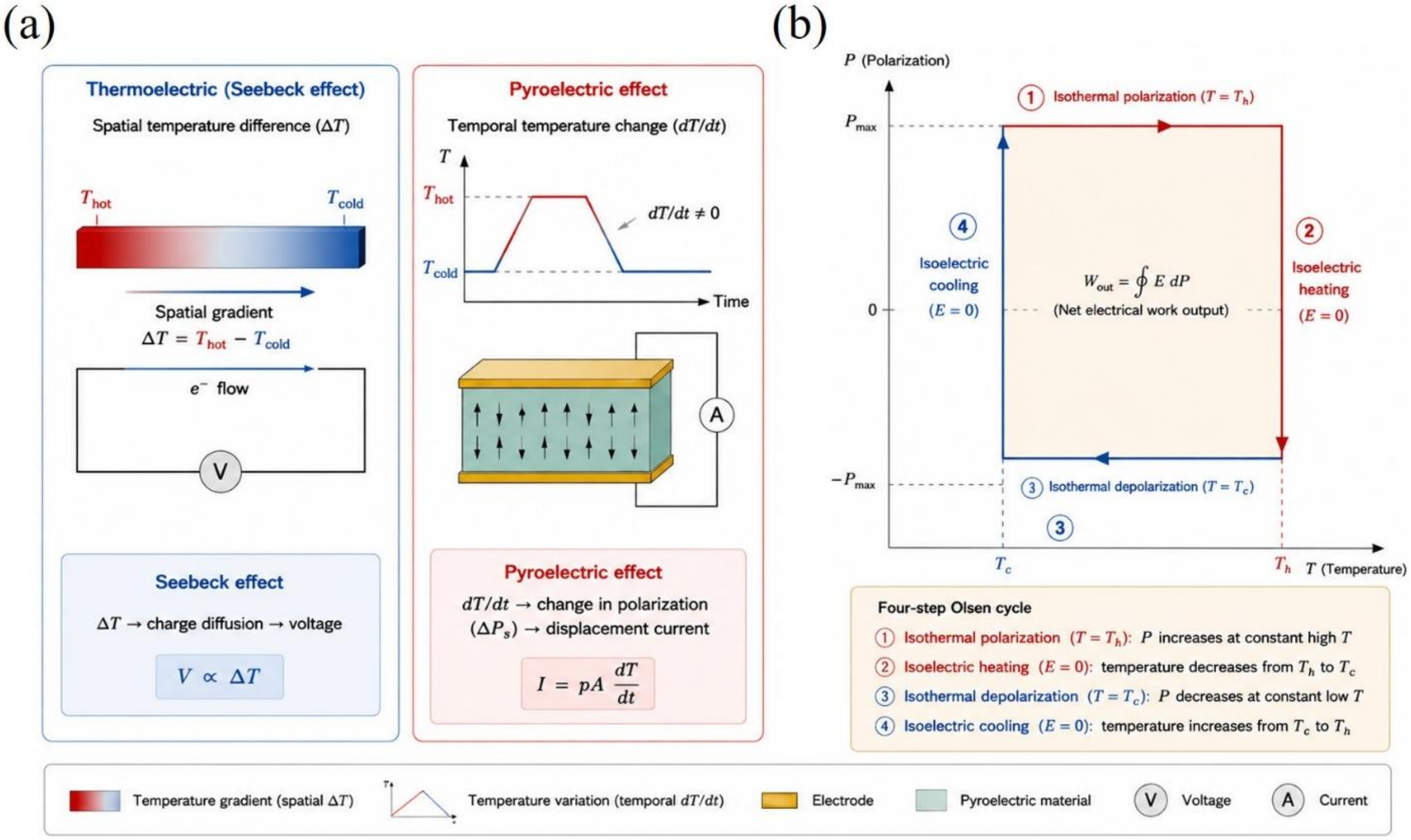

**Figure 10. Fundamental mechanisms of pyroelectric energy harvesting.** (a) Distinction between thermoelectric energy harvesting driven by a spatial temperature difference ($\Delta T$) through the Seebeck effect and pyroelectric energy harvesting driven by temporal temperature variations ($dT/dt$) through temperature-dependent polarization [76,77,85]. (b) Olsen thermodynamic cycle showing the four sequential processes of isothermal polarization, isoelectric heating, isothermal depolarization and isoelectric cooling for pyroelectric energy conversion [79].

### 4.2. The Olsen Cycle and Thermal Switching

A major strategy for overcoming the low power output associated with slow thermal fluctuations is to actively modulate temperature and electric field through thermodynamic cycles. The Olsen cycle, comprising two isothermal field-switching and two isoelectric heating/cooling processes, provides an effective route for increasing the electrical energy extracted per thermal cycle [79–81,83]. The material is polarized at low temperature, heated under a constant high field, depolarized at elevated temperature, and subsequently cooled under a constant low field to restore the initial state.

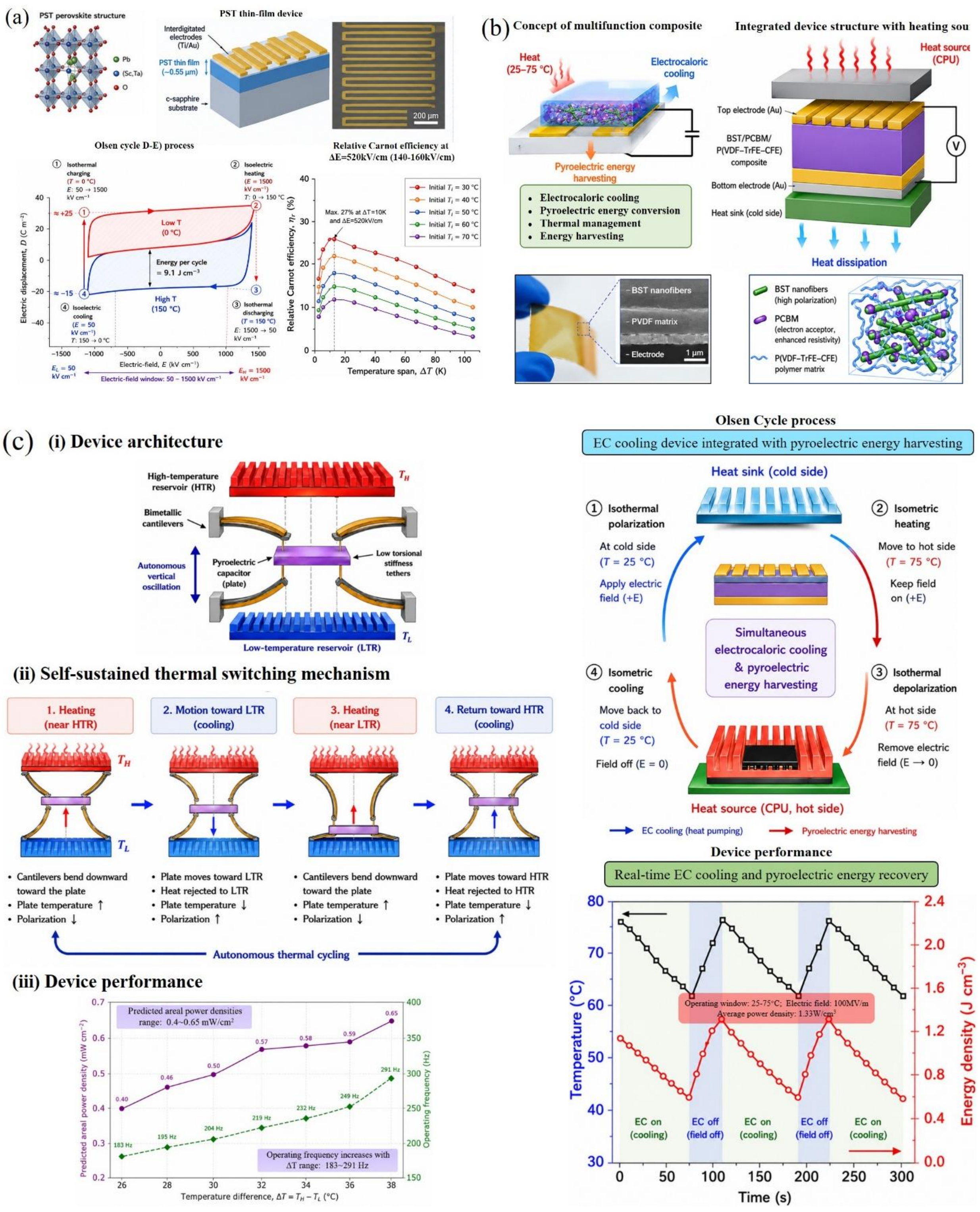

**Figure 11. Advanced strategies for high-performance pyroelectric energy harvesting.** (a) High-field Olsen-cycle harvesting using a $Pb(Sc_{1/2}Ta_{1/2})O_3$ (PST) thin film with interdigitated electrodes, achieving a calculated energy density of 9.1 J/cm³ per cycle between 0 and 150 °C and a relative efficiency of up to 27% of the Carnot limit under a 10 K temperature span [80]. (b) Integrated electrocaloric cooling and pyroelectric energy harvesting using a BST/PCBM/P(VDF-TrFE-CFE) ferroelectric polymer composite,

yielding 1.33 $J/cm^3$ over 25–75 °C while enabling simultaneous thermal management and energy recovery [81]. (c) Self-sustained thermal cycling driven by thermally actuated bimetallic cantilevers, enabling contact-free oscillation of a pyroelectric capacitor between hot and cold reservoirs. Modelling predicts an areal power density of 0.4–0.65 $mW/cm^2$ and an operating frequency of 183–291 Hz for reservoir temperature differences of 26–38 °C [82].

Recent advances have focused on high-field ferroelectrics and multifunctional architectures to increase cycle energy while reducing thermal and electrical losses. A $Pb(Sc_{1/2}Ta_{1/2})O_3$ (PST) thin-film harvester achieved a calculated energy density of 9.1 $Jcm^3$ per cycle between 0 and 150 °C [80] (Fig. 11(a)). Under a smaller 10 K temperature span and different electrical conditions, a relative efficiency of up to 27% of the Carnot limit was reported [80]. Meanwhile, integration of electrocaloric cooling with a BST/PCBM/P(VDF-TrFE-CFE) ferroelectric polymer composite enabled simultaneous thermal management and energy recovery, yielding 1.33 $J/cm^3$ over 25–75 °C [81] (Fig. 11(b)). However, maximizing energy per cycle does not necessarily maximize practical power output, which depends jointly on the harvested energy per cycle and thermal-cycling frequency [77]. Because heat transfer constrains the achievable cycling rate, self-sustained thermal switching has been explored to enhance thermal cycling without continuous external actuation. In a bimetallic pyroelectric harvester, thermally actuated cantilevers drive contact-free oscillation of the pyroelectric capacitor between hot and cold reservoirs [82]. As showing in Fig. 11(c), reduced-order modelling predicts that increasing the reservoir temperature difference from 26 to 38 °C raises the operating frequency from 183 to 291 Hz, with predicted areal power densities of approximately 0.4–0.65 $mW/cm^2$ [82]. These developments highlight a shift from maximizing energy extracted per thermal cycle toward system-level co-optimization of energy density, heat transfer, cycling frequency and conversion efficiency for practical pyroelectric harvesting [77,83].

## 5. Hybrid and Multi-Source Energy Harvesters: The Synergy Imperative

### 5.1. The Rationale for Hybridization

No single energy-harvesting modality is universally applicable. Piezoelectric and triboelectric harvesters depend on mechanical excitation, photovoltaic cells require illumination, and thermoelectric generators rely on sustained temperature gradients. Moreover, mechanical, optical, thermal and radio-frequency sources differ substantially in availability, intermittency, power density and electrical characteristics. Hybridization therefore provides a practical route toward more robust energy autonomy by integrating complementary transduction mechanisms within a common device or system. Hybridization can occur at different levels. Independent outputs may be electrically combined or temporally complemented under changing environmental conditions, whereas shared-footprint integration allows multiple transducers to share active area, electrodes or structural components. Stronger functional coupling arises when one transduction process modifies another through electrical, interfacial, thermal or mechanical interactions. These modes are not mutually exclusive and may coexist within a single hybrid architecture. Importantly, hybridization does not inherently imply energetic synergy. Integration may improve energy availability or reduce footprint without generating more useful energy than the constituent harvesters independently. Genuine synergy should therefore be distinguished from simple output addition and assessed at a common system boundary under matched input conditions, with interface and control losses considered where relevant.

Figure 12 illustrates the progression of hybrid energy harvesting from device-level integration to functional coupling and, ultimately, system-level energy management. Shared-footprint PV–TENG devices combine solar and raindrop harvesting [88], whereas triboelectric–photovoltaic functional coupling exploits interfacial charge interaction to regulate carrier separation and promote cooperative light–mechanical conversion [55,89,90]. Photothermally activated piezo–pyroelectric energy harvesting and asymmetric thermo–mechanical coupling further exemplify multiphysics interactions [92,93]. At the system level, heterogeneous PV, TEG, piezoelectric and triboelectric sources require source-specific energy extraction, impedance management and power conversion before their outputs can be efficiently combined and delivered to energy storage or a common load [96–99]. Therefore, the performance of hybrid harvesters should ultimately be evaluated by the net useful energy delivered across the system boundary, rather than by isolated voltage, current or peak-power metrics. The representative architectures and coupling strategies summarized in Fig. 12 are examined in greater detail in Sections 5.2–5.4.

**(a) Shared-footprint PV-TENG for simultaneous solar and raindrop harvesting**

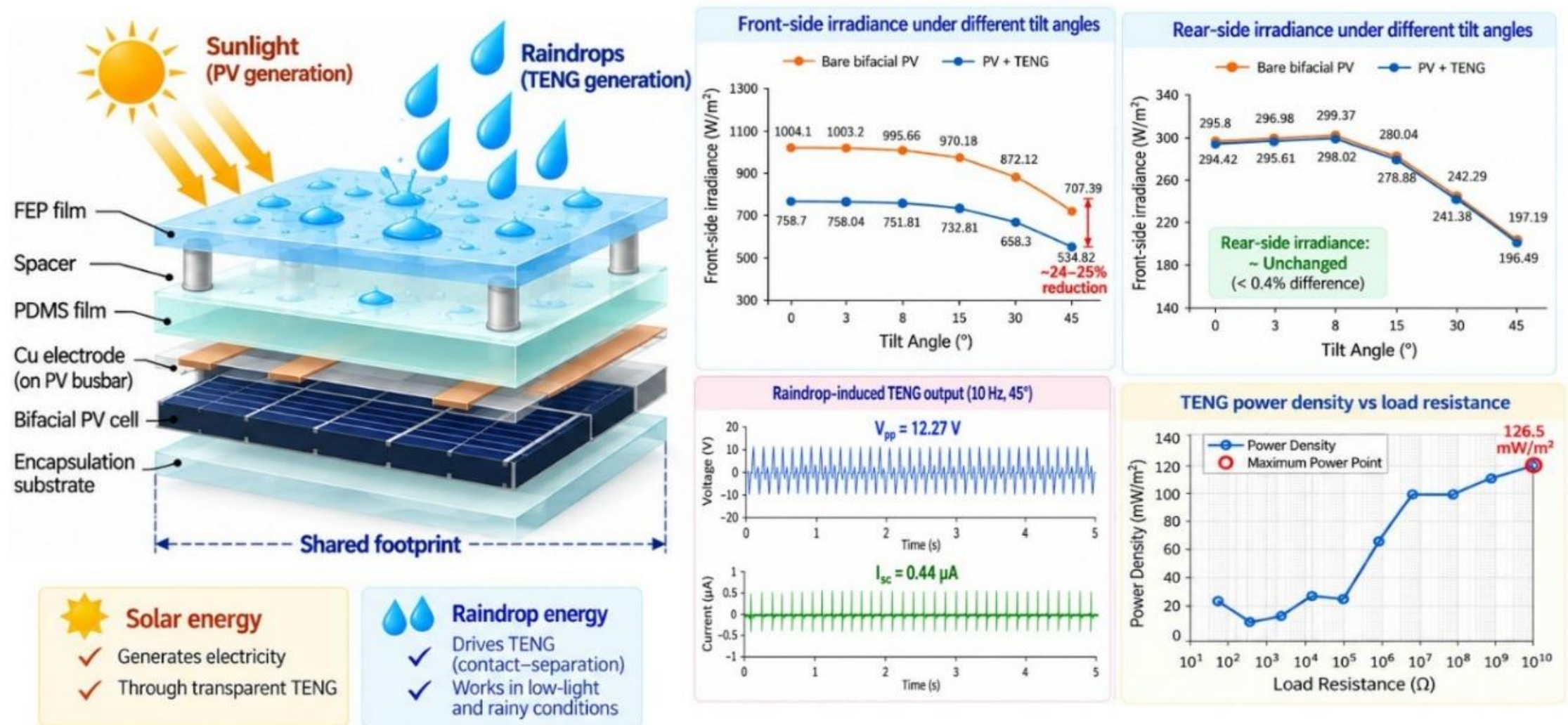

**(b) Triboelectric-photovoltaic function coupling**

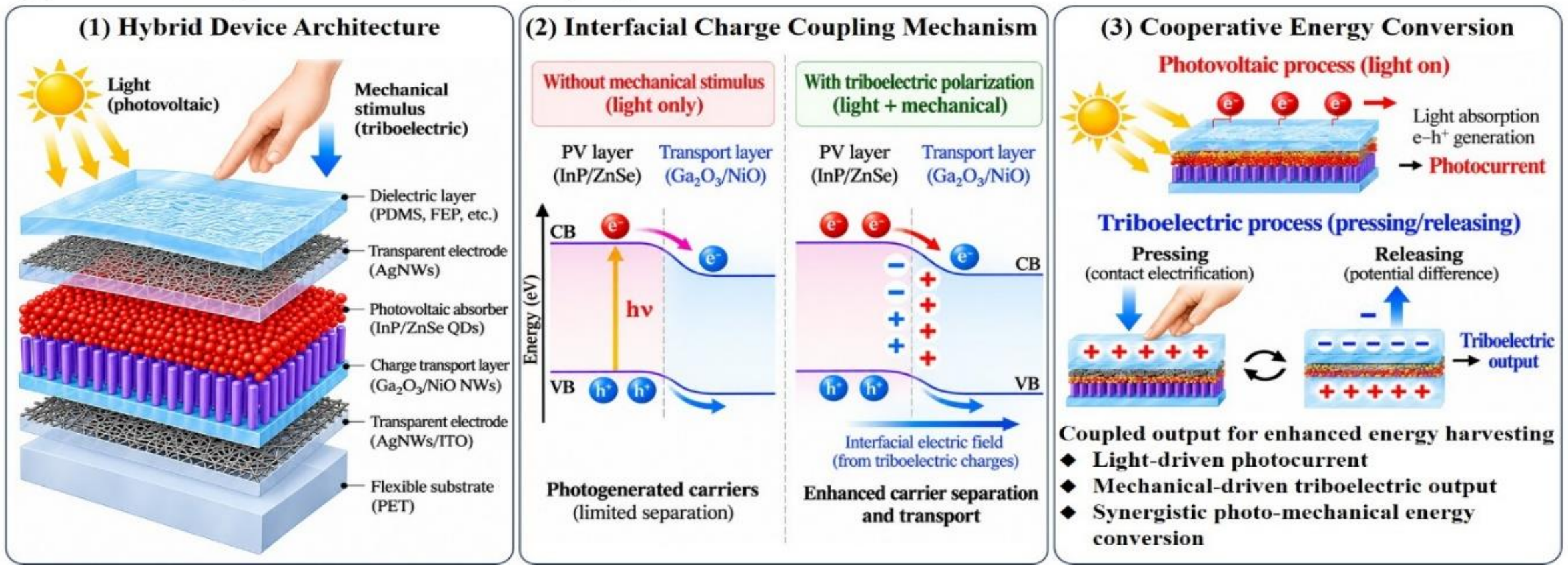

(c) Photothermally-activated piezo-pyroelectric multifunctional hybrid energy

(d) Asymmetric Thermo-Mechanical coupling

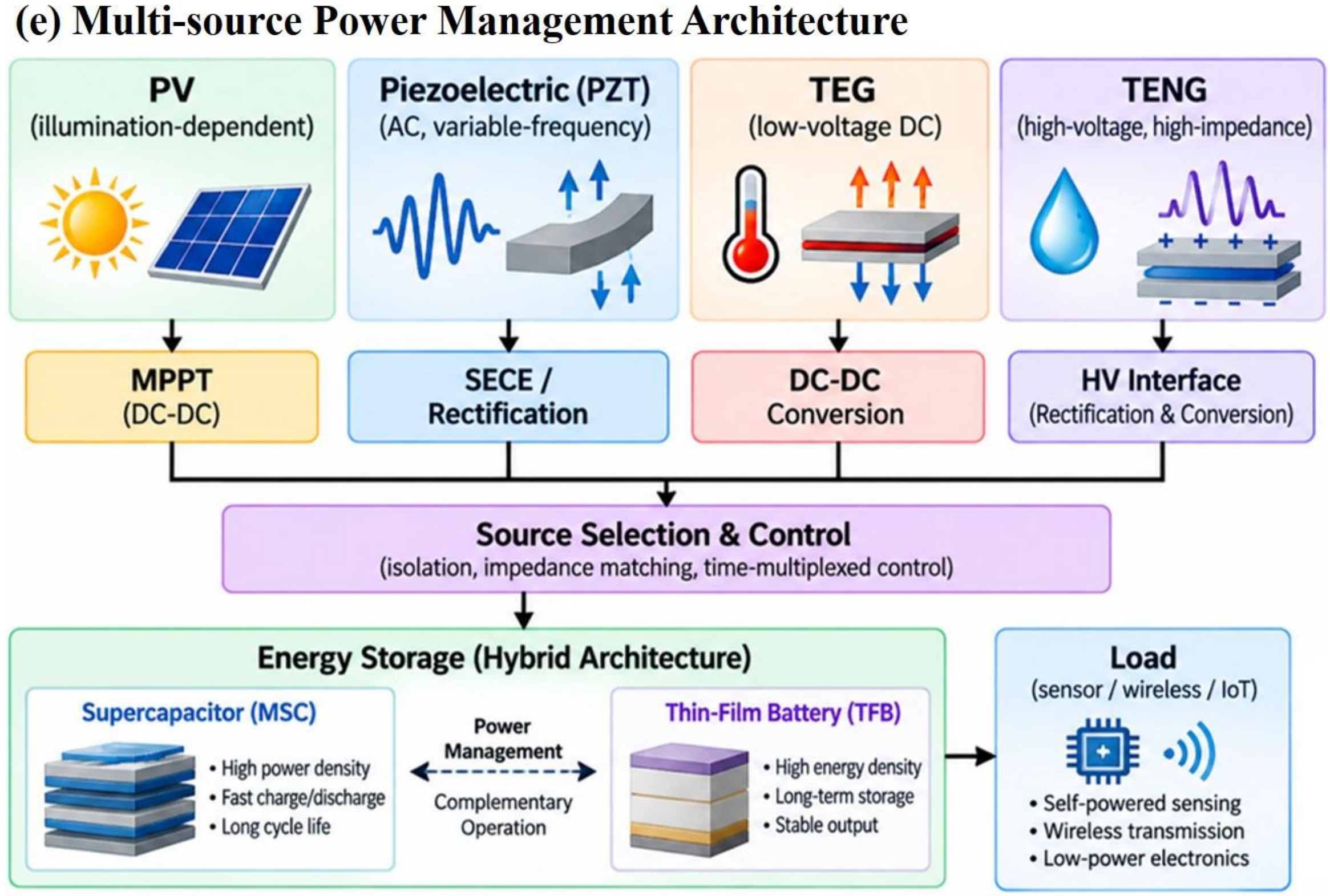

**Figure 12. Hybrid and multi-source energy-harvesting architectures.** (a) Shared-footprint PV–TENG integrating a transparent FEP/PDMS contact–separation TENG with a bifacial PV cell for simultaneous solar and raindrop harvesting; integration reduces front-side irradiance by ~24–25%, while the TENG delivers 12.27 V and ~126.5 mW/m² [88]. (b) Triboelectric–photovoltaic functional coupling through dielectric, field and interfacial charge engineering, enabling enhanced carrier separation and cooperative light–mechanical energy conversion [55,89,90] (c) Photothermally activated P(VDF-TrFE)/PEAn piezo–

pyroelectric harvester, in which solar-induced thermal cycling couples pyroelectric and piezoelectric outputs, yielding up to a 3.26-fold voltage enhancement during cooling [92]. (d) Asymmetric thermo–mechanical coupling in a lead-free BCT–BST pyro–piezoelectric nanogenerator, with current suppressed by 41.4% during heating with force but enhanced by 13.4% during cooling with force [93]. (e) Multi-source power management using source-specific extraction, MPPT, isolation, and shared conversion to interface heterogeneous PV, TEG, piezoelectric, and triboelectric harvesters [96–99].

### 5.2. PV–TENG Shared-Footprint and Coupled Architectures

PV–TENG systems illustrate the evolution from shared-footprint integration toward mechanism-level coupling. Early flexible architectures integrated a transparent single-electrode TENG with an organic solar cell on the same thin-film platform, enabling independent or simultaneous harvesting of solar and mechanical energy [95]. This integration introduces competing optical and electrical requirements: increasing Ag-nanowire coverage improves triboelectric performance but reduces optical transmittance, with ~85% selected as a practical compromise. Under strong illumination, the photovoltaic contribution dominates, whereas under weak indoor light the triboelectric contribution becomes increasingly important, demonstrating resource complementarity rather than intrinsic energetic synergy [95]. Recent designs increasingly exploit interactions between the constituent mechanisms. In a flexible $CsPbBr_3/CaF_2$ hybrid nanogenerator, fluoride-mediated dielectric regulation passivated defects, enhanced dielectric response and suppressed carrier trapping and recombination, yielding a maximum current of 9.7 μA under illumination, ~25-fold higher than the control condition [89]. Energy-band engineering and interfacial charge coupling have further enhanced carrier separation and cooperative light–mechanical energy conversion in integrated triboelectric–photovoltaic platforms [55,90], marking a transition from physical coexistence toward active cross-mechanism regulation (Fig. 12(b). Shared-footprint integration can nevertheless impose performance penalties. In a 2026 bifacial PV–contact-separation TENG, transparent FEP/PDMS layers enabled simultaneous solar and raindrop harvesting but reduced front-side irradiance by ~24–25% [88] (Fig. 12(a)). The TENG generated 12.27 V and a maximum power density of ~126.5 mW/m² under controlled dripping, yet contributed substantially less energy than the PV subsystem. Hybrid performance must therefore be evaluated against optical, electrical and interfacial penalties rather than inferred from simultaneous operation alone.

### 5.3. The "All-in-One" Thermal-Mechanical Harvester

Integrating thermal and mechanical transduction within a common active material provides a compact route to multifunctional energy harvesting. Earlier studies established the coexistence of pyroelectric and piezoelectric responses in miniaturized ferroelectric harvesters [91]. More recently, a flexible P(VDF-TrFE)-based piezo–pyroelectric harvester incorporating photothermal poly(2-ethylaniline) nanoparticles combined solar-induced heating with mechanical energy conversion [92] (Fig. 12(c)). During cooling, the output voltage increased by up to 3.26-fold relative to piezoelectric-only harvesting, while the same platform supported motion sensing and thermal therapy. Crucially, thermal and mechanical contributions are not necessarily additive. As evidenced by the state-dependent response shown in Fig.12(d), Zhao et al. reported on a lead-free BCT–BST pyro–piezoelectric nanogenerator that exhibit a piezoelectric current of 54.4 nA, whereas simultaneous heating and mechanical loading reduced the current by 41.4%; by contrast, cooling under mechanical loading enhanced it by 13.4% [93]. This asymmetric response was attributed to modulation of band bending through coupled pyroelectric and piezoelectric effects. Under optimized cooling–force conditions, the device charged a 4.7 μF capacitor to 3 V within 46 s and operated stably for 9 h. Overall, these results highlight a defining feature of multiphysics functional coupling: interactions between transduction mechanisms can be either constructive or destructive depending on the operating state. Hybrid performance should therefore be evaluated under explicitly defined thermal and mechanical conditions, rather than inferred from the coexistence of multiple transduction responses.

### 5.4. The Critical Bottleneck: MPPT and Power Management for Hybrids

As hybrid harvesters become more complex, the principal bottleneck increasingly shifts from transduction to power management. PV cells require illumination-dependent maximum-power-point tracking (MPPT), TEGs provide low-voltage DC, piezoelectric harvesters produce alternating outputs, and TENGs typically generate high-voltage, high-impedance pulses. Effective hybrid interfaces therefore require source-specific extraction, rectification, impedance matching and isolation before heterogeneous outputs can be combined downstream (Fig. 12(e)). Recent circuits increasingly share conversion hardware while retaining source-specific control. Zhu et al. used multisource interface to harvest vibration and light from piezoelectric and photovoltaic transducers using peak-voltage detection for the piezoelectric inputs and an approximately $0.7V_{OC}$ photovoltaic MPPT criterion [96]. The circuit delivered 2.8 mW, achieved a maximum PV extraction efficiency of ~75%, and produced 4.59-fold higher output power than operation without the photovoltaic contribution. A related self-powered interface combined synchronous electric charge extraction with photovoltaic MPPT through a compact flyback topology, enabling simultaneous extraction from heterogeneous sources [97]. Scaling to multiple independently varying inputs further increases control complexity. Chen et al. reported on multi-input piezoelectric MPPT architecture employing an adaptive switch network and single-inductor interface for simultaneous and individual voltage regulation [98]. At the system level, von Kluge et al. demonstrated a PV–TEG–piezoelectric platform combining MPPT with dynamic source switching able to harvest 28–35% more energy than the best-performing standalone source under fluctuating irradiance, temperature gradients and vibration [99]. Nevertheless, sensing, switching, leakage and control consume harvested energy and can perturb individual sources from their optimal operating points. Hybrid systems should therefore be benchmarked by the net useful energy delivered to a common load or storage element under realistic, time-varying inputs, rather than by simultaneous peak outputs or transducer-level metrics alone. The representative hybrid architectures discussed above are compared in Table 2 in terms of their constituent harvesters, integration strategies, synergy mechanisms, reported performance gains, target applications, and remaining challenges. Importantly, the reported gains arise from different levels of integration, ranging from shared-footprint resource complementarity to mechanism-level coupling and source-aware electrical management. They should therefore be interpreted within their respective operating conditions rather than as directly interchangeable performance metrics.

**Table 2. Hybrid Harvester Architectures: Performance and Synergy**

| Hybrid Combination | Architecture | Synergy Mechanism | Performance Gain | Application | Key Challenge |
| --- | --- | --- | --- | --- | --- |
| **PV + TENG** | Transparent TENG + bifacial PV [88] | Complementary solar/rain harvesting [55,88–90] | TENG: 12.27 V, ~126.5 mW m$^{-2}$; irradiance −24–25% [88] | All-weather sensing | Optical loss vs. triboelectric output |
| **PV + TEG + Piezoelectric** | Three-source platform + dynamic switching [99] | Source complementarity | 28–35% more energy than best standalone source [99] | Autonomous IoT / SHM | Control overhead |
| **Piezoelectric + Pyroelectric** | Multifunctional piezo–pyro platform [92,93] | Thermo-mechanical coupling | Up to 3.26× voltage [92]; +13.4% current under cooling + force [93] | Wearable sensing | Constructive/destructive coupling |
| **Triboelectric + Pyroelectric + RF** | Multi-source concept | Resource complementarity | No matched quantitative benchmark | Multi-source IoT | Impedance/interface mismatch |
| **Piezoelectric + Electromagnetic** | Hybrid flow harvester [131] | Complementary PE/EM transduction | 831.7 µW AC; 680 µW conditioned DC [131] | In-pipe monitoring | Hydraulic penalty / fouling |
| **PV + RF** | Heterogeneous-source integration | Optical/RF complementarity | No matched quantitative benchmark | Indoor IoT | Source-specific power management |

Collectively, these examples show that the value of hybridization lies not simply in combining multiple transducers, but in increasing source availability and net usable energy while minimizing the optical, mechanical, electrical, and control penalties introduced by integration.

## 6. Radiative Energy Harvesting: Indoor Photovoltaics, Ambient RF, and Backscatter

Radiative energy harvesting exploits ambient electromagnetic radiation across two distinct regimes: optical illumination and radio-frequency (RF) fields. Although sharing a common electromagnetic origin, their resource characteristics and conversion pathways differ fundamentally. Indoor photovoltaics operate under low-intensity, spectrally confined artificial illumination, shifting the design priority from broadband solar conversion toward spectrum–absorber matching, bandgap optimization, and suppression of low-light losses. Ambient RF harvesting, by contrast, must capture weak, spatially and spectrally variable electromagnetic fields before impedance matching and rectification, making multiband capture and efficient low-power conversion central challenges. Meaningful performance comparison therefore requires clearly defined reference planes along each conversion chain. For indoor PV, illuminance, spectrum, incident optical irradiance, active area, and electrical output should be distinguished; similarly, for RF harvesting, incident field strength, captured RF power, rectifier input, and DC output should be clearly differentiated. Fig. 6 summarizes these complementary radiative energy-harvesting pathways: indoor PV progresses from the distinct indoor spectral environment to bandgap and molecular engineering, whereas RF systems progress from multiband energy capture to backscatter communication and passive wake-up. Together, these approaches illustrate two complementary routes toward energy-autonomous IoT: increasing the amount of ambient energy that can be effectively harvested and reducing the energy required for sensing and communication. The underlying material, device, and system-level strategies summarized in Fig. 13 are examined in greater detail in the following sections.

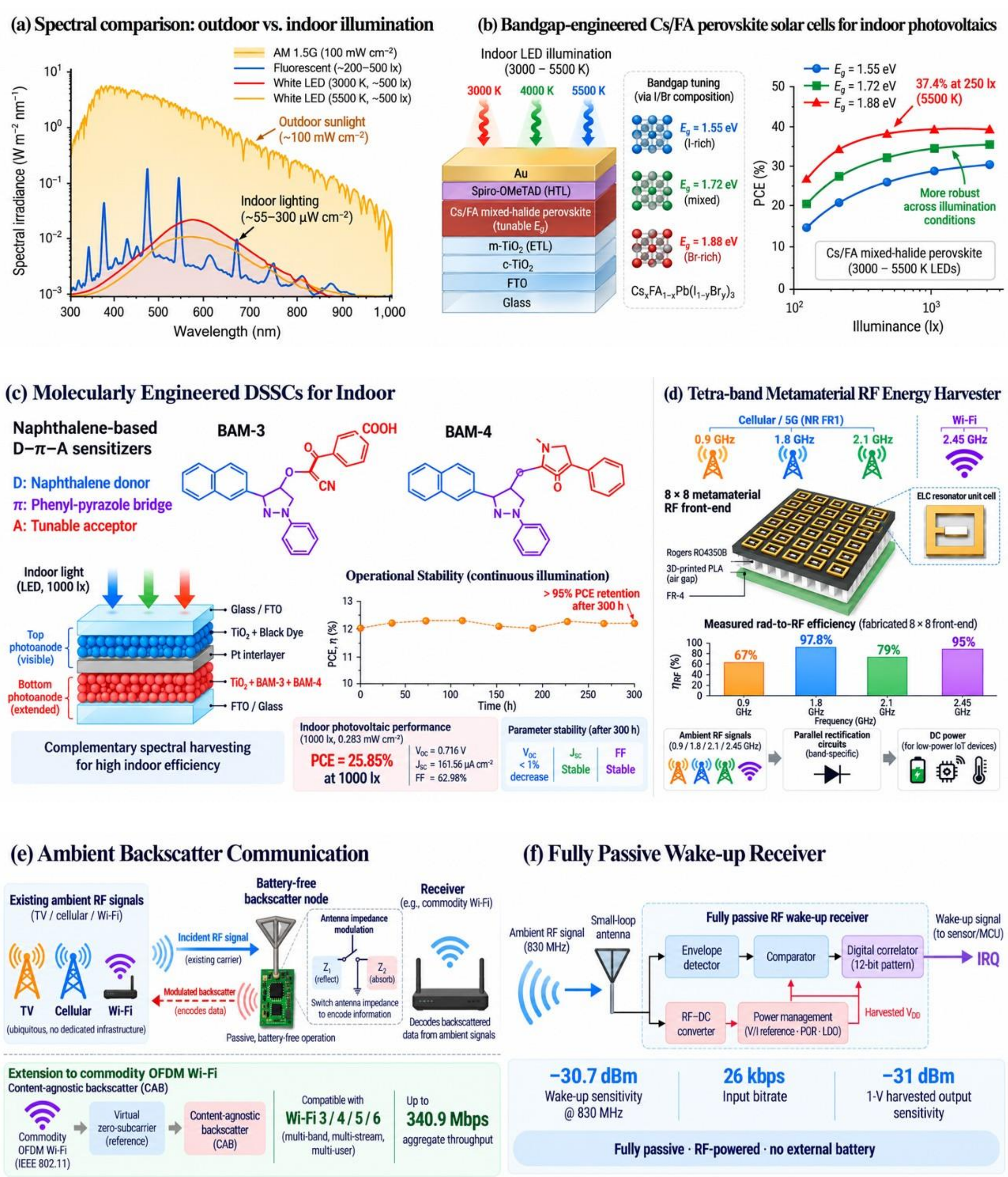

**Figure 13. Radiative energy harvesting: indoor photovoltaics, ambient RF, and backscatter.** (a) Spectral comparison of AM1.5G sunlight with lower-irradiance, spectrally confined indoor fluorescent and LED illumination, illustrating the distinct spectral environment encountered by indoor photovoltaics. (b) Bandgap-engineered Cs/FA mixed-halide perovskites (1.55–1.88 eV), with a 1.88-eV device reaching 37.4% PCE at 250 lx under a 5500 K LED [100]. (c) Molecular engineering of naphthalene-based D–π–A sensitizers and tandem co-sensitization in indoor DSSCs, achieving 25.85% PCE at 1000 lx, with >95% PCE retention after 300 h of continuous-illumination stability testing [101]. (d) Tetra-band ELC metamaterial RF harvester operating at 0.9, 1.8, 2.1 and 2.45 GHz, enabling

multiband electromagnetic capture across cellular and Wi-Fi bands [102]. (e) Ambient backscatter enables battery-free communication by modulating antenna impedance to encode information onto existing RF signals, with recent implementations extending this principle to commodity OFDM Wi-Fi [103,104]. (f) Fully passive RF wake-up receiver integrating RF energy harvesting and wake-up detection, with −30.7 dBm sensitivity at 830 MHz and a 26-kbps data rate [105].

## 6.1. Indoor Photovoltaics (PV): The Shift from Sun to Fluorescent/LED

Conventional silicon solar cells optimized for AM1.5G sunlight are not ideally suited to spectrally confined indoor illumination. As illustrated in Fig. 13(a), typical indoor lighting (200–1000 lx; ~55–300 μW/cm², spectrum-dependent) is far weaker than AM1.5G sunlight (100 mW/cm²), reducing photocarrier density and $V_{OC}$while increasing the relative importance of recombination and spectral mismatch. Indoor PVs therefore require spectrum–absorber matching, bandgap optimization, and low-light loss suppression. Because lux does not uniquely represent optical irradiance, rigorous comparisons should additionally report illumination spectrum, calibrated irradiance, active area, and operating conditions [106]. PSCs and DSSCs are particularly attractive because their spectral responses can be tailored through composition/bandgap and molecular engineering, respectively.

**Bandgap Tuning via Cesium/Formamidinium Mixing:** The composition and bandgap of PSCs can be tailored to match artificial-light spectra. Modeling of Cs/FA mixed-halide PSCs predicted a maximum PCE of 28.38% and a power density of 106.25 μW/cm² under white LEDs [107], while triple-cation devices spanning ~1.60–1.77 eV achieved up to 34.0% PCE at 870 lx [108]. More recently, Cs/FA mixed-halide absorbers with bandgaps of 1.55, 1.72, and 1.88 eV were evaluated under 3000–5500 K LEDs at 250–1000 lx [100]. As summarized in Fig. 13(b), the 1.72-eV device reached 35.04% PCE at 1000 lx and 36.6% at 250 lx, whereas the 1.88-eV device achieved 37.4% at 250 lx under a 5500 K LED. These results show that indoor performance is governed by spectrum-dependent optimization of photon absorption, photovoltage, and low-light recombination rather than by bandgap narrowing alone.

**Dye-Sensitized Solar Cells (DSSCs) with Molecular Engineering:** DSSCs provide a complementary route in which spectral response and interfacial charge transfer are tailored through sensitizer chemistry. Donor, π-linker, and acceptor units can be engineered to regulate absorption, electron injection, aggregation, and recombination [109]. Fig. 13(c) describes this strategy using naphthalene-based D–π–A sensitizers and tandem spectral harvesting, achieving 25.85% PCE at 1000 lx and >95% PCE retention after 300 h of continuous illumination [101]. Indoline-donor engineering further yielded 31.34% PCE under fluorescent illumination at 2500 lx [110]. Thus, PSCs and DSSCs provide complementary routes to indoor spectral matching through bandgap and molecular engineering, respectively. Device-level efficiency alone, however, does not ensure energy-autonomous operation. Practical output is further constrained by active area, illumination conditions, PMIC losses, storage leakage, and load duty cycle. A storage-assisted LoRaWAN sensor node, for example, operated from an amorphous-Si photovoltaic source at ~200 lx with a reported operating point of 3.0 V and 45.7 μA [1]. Indoor-PV development must therefore balance spectral conversion and stability with power management, storage, and system-level energy demand. The major indoor-photovoltaic platforms considered in this review are compared in Table 3. Their performance reflects fundamentally different absorber chemistries and spectral-matching strategies, and the reported efficiencies must be interpreted together with illuminance, illumination spectrum, color temperature, and measurement conditions. Accordingly, the table is intended to identify technology-specific strengths and limitations rather than to establish a universal efficiency ranking.

**Table 3. Comparison of Indoor Photovoltaic Technologies**

| Parameter | Amorphous Silicon (a-Si) | Perovskite (PSC) | Dye-Sensitized (DSSC) | Organic PV (OPV) |
|---|---|---|---|---|
| **Best PCE @ 500 lux (2024–2026)** | No matched 500-lx PCE benchmark; 3.0 V, 45.7 µA at ~200 lx demonstrated [1] | Up to 34.6% at 500 lx, 1.72-eV Cs/FA PSC [100] | No matched 500-lx benchmark | Not quantitatively benchmarked |
| **Best PCE @ 1000 lux (2024–2026)** | Not quantitatively benchmarked | 35.04% at 1000 lx, 1.72-eV Cs/FA PSC [100] | 25.85% at 1000 lx [101] | Not quantitatively benchmarked |
| **Optimal Wavelength Range** | Broad visible response | Bandgap-/spectrum-tunable [100,107,108] | Dye-specific spectral tuning [101,110] | Molecularly tunable |
| **Low-Light Stability** | System operation demonstrated at ~200 lx [1] | Long-term indoor operation demonstrated for selected devices [100] | >95% retention after 300 h [101] | Not quantitatively benchmarked |
| **Lead Content** | None | Pb-containing in cited Cs/FA devices [100] | None | None |
| **Manufacturing Cost** | Established thin-film process; architecture-dependent | Potentially scalable solution processing | Dye/electrolyte/process-dependent | Process- and material-dependent |
| **Form Factor / Flexibility** | Glass or flexible thin-film | Thin-film; flexible architectures possible | Thin-film / mesoscopic; flexible possible | Thin-film; high flexibility potential |
| **Commercial Readiness** | Established | Emerging / research-to-pilot | Emerging / niche | Emerging |
| **Best Application** | Indoor IoT / storage-assisted sensors [1] | High-efficiency indoor PV [100,107,108] | Spectrum-tailored indoor PV [101,110] | Flexible / lightweight indoor PV |

Whereas indoor photovoltaics exploit relatively concentrated optical radiation through semiconductor absorption, ambient RF harvesting operates under substantially weaker and more spatially variable electromagnetic fields, shifting the dominant challenges toward electromagnetic capture, impedance matching, and weak-signal rectification.

## 6.2. RF Energy Harvesting: Metamaterials and Ambient Backscatter

Ambient RF energy from cellular, Wi-Fi, television, and other wireless infrastructure provide a widely distributed but highly environment-dependent resource. Unlike dedicated wireless power transfer, ambient harvesting cannot control transmitter power, frequency, polarization, or propagation distance, and available RF power varies strongly with location and frequency. These constraints shift the design priority toward multiband electromagnetic capture, efficient rectification at weak input power, and communication architectures that minimize node-level energy consumption.

**Metamaterial-Based Broadband Rectennas:** Conventional RF harvesters are typically optimized for one or a few resonances, whereas ambient RF energy is distributed across multiple communication bands. Metamaterial architectures provide a route to multiband capture through engineered subwavelength resonators. Fig. 13(d) highlights an 8 × 8 tetra-band ELC metamaterial harvester operating at 0.9, 1.8, 2.1, and 2.45 GHz [102], with measured radiation-to-RF efficiencies of approximately 67%, 97.8%, 79%, and 95%, respectively. A unit cell generated up to 562 μW at an incident RF power density of 40 μW/cm². Complementarily, a meta-lens-assisted harvester enhanced received RF power by >10 dB over ~2.9–3.63 GHz and maintained >30% total conversion efficiency near −20 dBm [111]. Importantly, broadband electromagnetic capture should not be equated with broadband RF-to-DC conversion unless matching, rectification, and DC output are characterized over the same operating range. At weak RF levels, rectification remains a critical bottleneck because conventional diode efficiency deteriorates as the available voltage decreases. Nanoscale spin-rectifier arrays, for example, operated from −62 to −20 dBm, achieving a zero-bias sensitivity of ~34,500 mV/mW and a peak conversion efficiency of 7.81%, with sensor operation demonstrated at −27 dBm [112]. These results emphasize that practical RF harvesting requires co-optimization of electromagnetic capture, impedance matching, and low-power rectification.

**Ambient Backscatter and Passive Wake-Up:** Rather than increasing harvested DC power alone, backscatter reduces communication demand by modulating antenna impedance and reusing an existing RF carrier instead of generating one locally. As presented in Fig. 13(e), ambient backscatter enables battery-free communication through controlled modulation of the reflected RF field [103]. More recently, content-agnostic backscatter extended this principle to commodity OFDM Wi-Fi, demonstrating compatibility with Wi-Fi 3/4/5/6 and aggregate throughput up to 340.9 Mbps [104]. Passive wake-up receivers further reduce standby consumption by activating the node only when an RF trigger is detected. Fig. 6f shows a fully passive architecture integrating RF–DC energy harvesting with envelope detection and digital correlation; the demonstrated receiver achieved −30.7 dBm sensitivity at 830 MHz and a 26-kbps input bitrate [105]. Together, Fig. 13(d)–(f) illustrates the complementary RF strategies of harvesting more and consuming less: multiband structures improve electromagnetic capture, low-power rectification converts weak RF inputs, while backscatter and passive wake-up reduce communication and standby energy. System performance therefore depends on the complete chain from ambient RF availability and antenna capture to matching, rectification, power management, and communication demand. These complementary RF strategies are compared in Table 4. Conventional antennas and metamaterial rectennas primarily address electromagnetic capture and energy conversion, whereas ambient backscatter reduces the energy required for communication by reusing existing RF carriers. The corresponding metrics therefore refer to different points along the RF source-to-load pathway and should not be interpreted as a single efficiency hierarchy.

**Table 4. RF Energy Harvesting Technologies Comparison**

| Parameter | Conventional Patch Antenna | Metamaterial Rectenna | Ambient Backscatter |
|---|---|---|---|
| **Frequency Coverage** | Narrow / application-specific | Multiband: 0.9, 1.8, 2.1 and 2.45 GHz [102]; 2.9–3.63 GHz meta-lens enhancement [111] | Ambient RF carriers; OFDM Wi-Fi demonstrated [103,104] |
| **Absorption Efficiency** | Architecture-dependent | Rad-to-RF: ~67%, 97.8%, 79%, 95% at 0.9/1.8/2.1/2.45 GHz [102] | N/A; reflection/modulation [103,104] |
| **Sensitivity (min. input power)** | Rectifier-/matching-dependent | Operation from −62 to −20 dBm [112] | −30.7 dBm at 830 MHz [105] |
| **Output Voltage (at 20 μW/cm²)** | Input-/topology-dependent | No matched benchmark retained | N/A |

| Power Consumption | Passive front end | Passive front end | Fully passive wake-up front end [105] |
|---|---|---|---|
| **Communication Range** | N/A | N/A | Carrier- and environment-dependent [103,104] |
| **Data Rate** | N/A | N/A | 340.9 Mbps aggregate [104]; 26 kbps wake-up [105] |
| **Antenna Form Factor** | Patch / PCB | Metamaterial / meta-lens [102,111] | Antenna / impedance modulator [103–105] |
| **Key Component** | Matching network / rectifier | Engineered resonator/metamaterial capture + rectification [102,111,112] | Impedance modulator / passive detector [103–105] |
| **Rectification Technology** | Diode-based RF–DC | Weak-input / multiband RF–DC [102,111,112] | Not required for backscatter; used for wake-up [105] |
| **Primary Use** | RF energy harvesting | Multiband RF harvesting | Communication / wake-up |
| **Limitation** | Narrowband / matching sensitivity | Fabrication and low-input conversion losses | Requires ambient carrier / suitable link |

Taken together, these approaches demonstrate that energy autonomy can be improved either by increasing the energy captured from the environment or by reducing the energy required by the load. In both cases, however, the harvested energy must ultimately be buffered, conditioned, and delivered efficiently, making energy storage and power management the next critical system-level bottlenecks.

## 7. The Unseen Bottleneck: Energy Storage and Power Management Integrated Circuits (PMICs)

Even a high-performance energy harvester cannot sustain an autonomous microsystem unless its intermittent and source-dependent output can be efficiently stored, conditioned, and delivered to the load. Harvesters may produce high-voltage pulses, millivolt-level outputs, or strongly fluctuating power, whereas sensors, microcontrollers, and wireless transmitters generally require regulated low-voltage power. Energy storage and power-management integrated circuits (PMICs) therefore provide the temporal and electrical bridge between harvested energy and useful operation: storage buffers energy over time, while PMICs govern startup, extraction, conversion, source isolation, and load delivery. Once sufficient upstream energy is available, losses within this storage–interface subsystem can become a dominant bottleneck. Fig. 14 therefore frames energy storage and power management as a coupled system problem, spanning intermittent-energy buffering, complementary electrochemical storage, cold-start operation, and source-aware extraction and conversion.

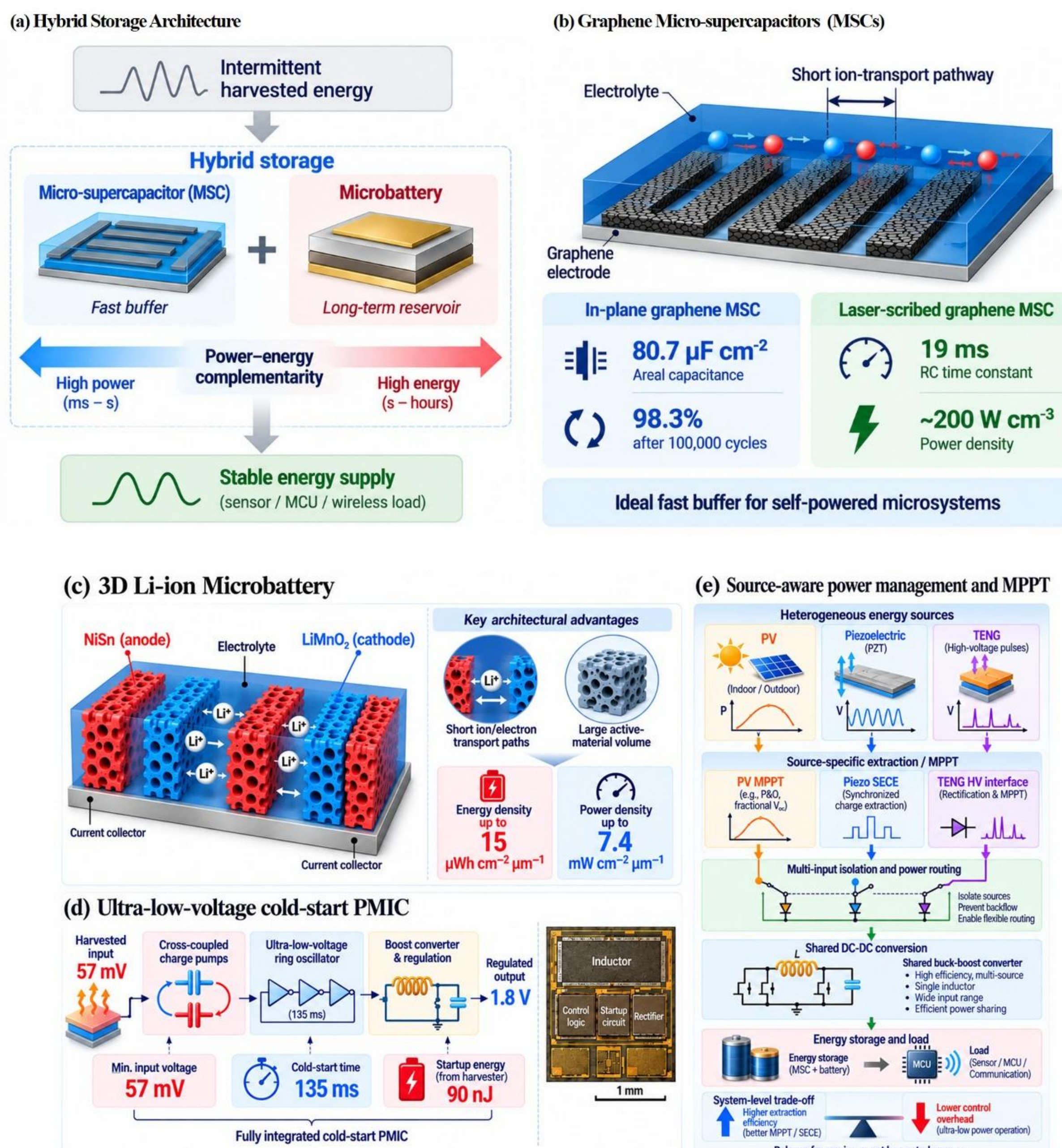

**Figure 14. Energy storage and power management: the unseen bottleneck.** (a) Hybrid storage architecture combining a micro-supercapacitor (MSC) for rapid buffering with a microbattery for longer-term energy storage [113,114,121]. (b) Graphene MSCs for rapid energy buffering: in-plane devices achieve an areal capacitance of 80.7 μF/cm² with 98.3% capacitance retention after 100,000 cycles [115], while laser-scribed MSCs exhibit an ~19 ms RC time constant and ~200 W $cm^{-3}$ power density [116]. (c) Interdigitated 3D Li-ion microbattery architecture combining short ion/electron transport paths with a large active-material volume, reaching energy and power densities of up to 15 μWh $cm^{-2}$ $μm^{-1}$ and 7.4 mW/cm² $μm^{-1}$, respectively [117]. (d) Fully integrated ultra-low-voltage PMIC enabling cold start of an inductive boost converter from 57 mV within 135 ms, while drawing approximately 90 nJ from the harvesting source [122]. (e) Source-aware power management combining source-specific PV MPPT, piezoelectric SECE, and high-voltage TENG extraction with input isolation and shared conversion for heterogeneous energy sources, while balancing extraction gains against control overhead [97,124–128].

### 7.1. Micro-Energy Storage: Supercapacitors vs. Thin-Film Batteries

Storage selection strongly influences the architecture and autonomy of an energy-harvesting microsystem. Nominal capacitance or battery capacity alone is insufficient; usable energy, equivalent series resistance (ESR), leakage, pulse capability, operating-voltage window, and degradation must be considered together [113,114]. Because MSC and microbattery performance is often normalized using different active-volume or footprint conventions, reported energy- and power-density values should be interpreted within their respective device definitions rather than as direct numerical rankings.

**Micro-Supercapacitors (MSCs):** MSCs are well suited to short-timescale buffering because of their high-power capability, rapid charge–discharge response, and cycling stability, making them attractive for intermittent outputs from TENGs and piezoelectric harvesters [113,114]. An in-plane graphene MSC achieved an areal capacitance of 80.7 $\mu F/cm^2$, volumetric energy and power densities of 2.5 $mWh/cm^3$ and 495 $W/cm^3$, respectively, while retaining 98.3% capacitance after 100,000 cycles [115]. Laser-scribed graphene MSCs further demonstrated an approximately 19 ms RC time constant and ~200 W $cm^{-3}$ power density [116]. These results highlight that capacitance, power capability, ESR, leakage, and cycling stability are distinct storage metrics.

For a capacitive buffer, usable energy is determined by its operating-voltage window:

$$E_{usable} = \frac{1}{2} C (V_{max}^2 - V_{min}^2)$$

while pulsed loading introduces an ESR-dependent voltage drop, $\Delta V = I R_{ESR}$. High capacitance alone therefore does not guarantee effective energy buffering, particularly when harvested energy must be accumulated over long dormant intervals.

**Thin-Film Solid-State Batteries (TFBs):** TFBs and micro batteries provide a complementary storage regime by retaining electrochemical energy over longer timescales. Their practical value depends on usable energy density, rate capability, internal resistance, voltage window, packaging, and cycling stability. Three-dimensional interdigitated lithium-ion micro batteries demonstrated energy densities up to 15 $\mu Wh/(cm^2 * \mu m)$ and power densities up to 7.4 $mW/(cm^2 * \mu m)$ [117], showing that microscale batteries are not inherently low-power devices. However, one characterized device retained only 64% of its initial energy after 15 cycles under a mixed-rate protocol, illustrating the trade-off between high power capability and cycling stability [117]. In comparison, rechargeable $Li/TiS_2$ solid-state thin-film cells with areal capacities of 35–100 $\mu Ah$ $cm^{-2}$ exceeded 1000 laboratory cycles, with selected cells exceeding 10,000 cycles under specified conditions [118]. Recent all-solid-state thin-film micro battery research has increasingly focused on electrode–electrolyte compatibility, ion-transport pathways, and three-dimensional or interface-engineered architectures to improve both integration and electrochemical performance [119,120].

Therefore, MSCs and microbatteries are complementary rather than mutually exclusive. As summarized in Fig. 14(a)–(c), hybrid storage can combine the rapid charge acceptance, high-power capability, and cycling durability of MSCs with the higher energy-storage capability of microbatteries. The complementary characteristics of these storage technologies are summarized in Table 5. MSCs and thin-film/microbatteries occupy different regions of the energy–power–timescale design space: MSCs favor rapid charge acceptance, high power delivery, and extensive cycling, whereas microbatteries provide greater retained energy over longer intervals. Hybrid storage seeks to exploit both regimes within a common energy-management architecture.

**Table 5. Energy Storage Technologies for Micro-Harvesters**

| Parameter | Micro-Supercapacitor (MSC) | Thin-Film Battery (TFB) | Hybrid (MSC + TFB) |
|---|---|---|---|
| **Energy Density** | 2.5 mWh $cm^{-3}$ demonstrated [115] | Up to 15 µWh $cm^{-2}$ $\mu m^{-1}$ [117] | Architecture-dependent [113,114,121] |
| **Power Density** | Up to 495 W $cm^{-3}$ [115]; ~200 W $cm^{-3}$ [116] | Up to 7.4 mW $cm^{-2}$ $\mu m^{-1}$ [117] | High transient capability [113,114,121] |
| **Charge/Discharge Rate** | Rapid | Chemistry- / architecture-dependent | Complementary rate capability |
| **Cycle Life** | ≥98.3% retention after 100,000 cycles [115] | >1,000 cycles; Li/$TiS_2$ selected cells >10,000 [118] | Cycling-regime-dependent [121] |
| **Self-Discharge Rate** | Higher; device-dependent [113–116] | Lower; device-dependent [117–120] | Timescale-dependent [121] |
| **ESR (Equivalent Series Resistance)** | Fast response; 19 ms RC demonstrated [116] | Chemistry- / architecture-dependent [117–120] | MSC supports pulse delivery |
| **Voltage Range** | Electrolyte-dependent; up to 2.5 V demonstrated [116] | Chemistry-dependent; 1.4–2.8 V for Li/$TiS_2$ [118] | Requires voltage-window coordination |
| **Recent Breakthrough (2024–2026)** | Microsystem integration [113,114] | Interface-engineered / 3D architectures [119,120] | Integrated storage management [113,114,121] |
| **Primary Role in Harvester** | Rapid energy buffer | Longer-term energy reservoir | Transient buffering + energy retention |
| **Cost** | Material / process-dependent | Chemistry / fabrication-dependent | Added integration complexity |
| **Form Factor** | Planar / flexible / on-chip [113–116] | Thin-film / 3D [117–120] | Integrated module |

The practical benefit of such hybridization, however, depends on whether the improved temporal matching between intermittent harvesters and pulsed loads outweighs the additional leakage, switching, sensing, and conversion losses introduced by the storage interface. This requirement places the PMIC at the center of the source–storage–load energy pathway.

## 7.2. PMIC Breakthroughs: Cold-Start, MPPT, and Ultra-Low Quiescent Current

Harvesting PMICs must operate across fundamentally different source conditions, ranging from millivolt-level thermoelectric outputs and illumination-dependent photovoltaic voltages to high-voltage, high-impedance piezoelectric and triboelectric pulses. PMIC performance is therefore source-specific, and cold-start voltage, tracking efficiency, conversion efficiency, quiescent consumption, and net delivered power should be evaluated separately.

**Sub-100 mV Cold-Start Capability:** Cold start remains a critical challenge for low-voltage energy harvesters because the PMIC must initially energize its oscillator, charge pump, and control circuitry before efficient steady-state conversion can begin. Bose *et al.* demonstrated an integrated electrical cold-start scheme that activates an inductive boost converter from 57 mV using cross-coupled complementary charge pumps and an ultra-low-voltage ring oscillator, achieving startup within 135 ms while drawing only 90 nJ from the harvesting source (Fig. 14(d)) [122]. Earlier thermoelectric interfaces achieved startup from approximately 35 mV using mechanical assistance before transitioning to regulated operation [123]. However, the minimum startup voltage alone does not determine practical cold-start performance. Startup energy and time, available source power, and the ability to transition autonomously into sustained operation must be considered together. This distinction is particularly important for thermoelectric and indoor-photovoltaic harvesters, where a low source voltage may coincide with severely limited available power.

**"Perturb-Observe with Hysteresis" (POH) MPPT:** Maximum-power-point tracking (MPPT) reduces source–interface mismatch by controlling the electrical operating point presented to the harvester. Conventional perturb-and-observe (P&O) approaches repeatedly perturb this point and can incur oscillation and control overhead around the maximum-power region. Hysteretic, event-driven, or intermittently sampled variants reduce unnecessary tracking activity by allowing operation within a prescribed window before the optimum point is re-evaluated, which is particularly attractive for microwatt-scale ambient sources. The relevant quantity is therefore the net benefit after controller overhead:

$$\mathrm{P}_{net,MPPT} = P_{harvested,MPPT} - P_{control}$$

Recent high-voltage PMIC development for TENGs has combined P&O-based maximum-power extraction with ultra-low-power voltage and current sensing, demonstrating that MPPT can be extended to the unusual high-voltage and high-impedance characteristics of triboelectric sources [124]. For low-light photovoltaics, an intermittently sampled interface operated over approximately 100–5000 lx while sampling the operating point once per minute and consuming an MPPT-control quiescent current of 8 μA [125]. Earlier multisource processors reported photovoltaic tracking efficiencies up to 96%, while peak power-conversion efficiencies differed substantially among photovoltaic, thermoelectric, and piezoelectric conversion paths [126]. These results emphasize that tracking efficiency, conversion efficiency, and quiescent consumption are distinct metrics. At very low source power, a slightly less accurate but intermittently activated MPPT can deliver more net energy than continuously active tracking if the reduction in controller consumption exceeds the energy lost away from the instantaneous maximum-power point. Recent ultra-low-quiescent-current energy-management circuits further address this constraint by minimizing always-on voltage references, comparators, and logic blocks [124].

**Synchronized Dual-Input Harvesting (No Parasitic Drain)**: Multisource harvesters introduce an additional PMIC challenge because individual sources vary independently in voltage, impedance, waveform, and availability. Direct electrical combination can force sources away from their optimum operating points or allow an active source to lose energy through an inactive path. Source-specific extraction, active isolation, and time-multiplexed conversion are therefore required for efficient multisource operation. A foundational photovoltaic–thermoelectric–piezoelectric processor accepted source voltages from approximately 20 mV to 5 V and used a shared-inductor switching network with source-specific maximum-power extraction [126]. More recently, a self-powered piezoelectric–photovoltaic interface integrated synchronized electric charge extraction (SECE), photovoltaic MPPT, and fast self-startup within a common interface, enabling simultaneous extraction from the two heterogeneous sources [97]. This architecture illustrates how source-specific extraction can be retained while reducing duplication of conversion hardware. Source isolation becomes particularly important when one environmental resource disappears. Reverse current through inactive rectification paths, switch leakage, sensing overhead, and quiescent consumption can gradually erode the small amount of stored ambient energy. Accordingly, “no parasitic drain” should be treated as a measurable system requirement rather than an assumed property of a dual-input topology. A particularly severe interface mismatch occurs for TENGs, whose high-voltage, low-current, and high-impedance outputs differ fundamentally from the low-voltage operating window of most storage elements and electronic loads. Recent work has therefore increasingly co-designed the TENG energy cycle and its power-management circuit rather than treating the PMIC as an independent downstream block. Gao et al. demonstrated a high-voltage/high-charge TENG energy cycle together with a synchronous power-management strategy designed to suppress transmission losses [127], while subsequent self-driven circuits have targeted efficient energy transfer even from relatively weak TENG outputs [128]. High transducer voltage alone is therefore not a sufficient indicator of usable system power; efficient charge extraction, impedance transformation, voltage conversion, and low-leakage switching determine how much of the generated energy ultimately reaches storage. Taken together, these studies establish a broader design principle for heterogeneous energy harvesting: source-specific extraction should be preserved upstream, whereas isolation, routing, and conversion hardware can be shared downstream when the resulting reduction in circuit duplication outweighs the associated switching, leakage, and control losses (Fig. 14(e)). Photovoltaic sources are naturally paired with MPPT, piezoelectric harvesters can benefit from synchronized charge-extraction strategies, and high-voltage, high-impedance TENG outputs require dedicated rectification and charge/voltage-management interfaces. Source isolation prevents inactive or mismatched inputs from parasitically loading active harvesters, while shared conversion can reduce duplicated hardware across multiple inputs. The relevant optimization target is therefore not peak extraction efficiency alone, but the net energy delivered to storage and the load after accounting for sensing, switching, leakage, and control overhead [97,124–128]. The principal PMIC requirements and representative advances discussed in this section are summarized in Table 6. Importantly, minimum cold-start voltage, startup energy and time, MPPT accuracy, conversion efficiency, quiescent consumption, source isolation, and leakage describe different aspects of interface performance. Their significance depends strongly on source voltage, impedance, available input power, and operating duty cycle; consequently, no single metric is sufficient to define a universally superior harvesting PMIC.

**Table 6. PMIC (Power Management IC) Specifications for Energy Harvesting**

| Parameter | Conventional PMIC (Pre-2024) | Representative Advanced PMICs | System-Level Advance |
|---|---|---|---|
| **Minimum Cold-Start Voltage** | Source- and architecture-dependent | 57 mV fully electrical startup; 135 ms; ~90 nJ [122] | Low-voltage autonomous startup |
| **MPPT Algorithm** | Continuous P&O / fixed-point tracking | High-voltage TENG P&O [124]; intermittent low-light PV sampling [125] | Source-aware, lower-overhead tracking |
| **MPPT Power Consumption** | Implementation-dependent | 8 µA MPPT-control quiescent current with once-per-minute sampling [125] | Reduced control overhead at low source power |
| **Quiescent Current** | Circuit- and duty-cycle-dependent | Ultra-low-overhead control emphasized [124,125] | Lower standby energy loss |
| **Peak Boost Efficiency** | Input- and topology-dependent | Tracking and conversion efficiencies reported separately [126] | Avoids conflating tracking with conversion efficiency |
| **Multi-Input Support** | Predominantly single-source | Source-specific extraction with shared conversion [97,126] | Heterogeneous-source integration |
| **Reverse Leakage Current** | Source- and switch-dependent | Input isolation to suppress parasitic drain [97,124–128] | Reduced loading of inactive sources |
| **Adaptive Sampling** | Fixed / continuous tracking | Intermittent sampling; once-per-minute PV sampling demonstrated [125] | Lower sensing and tracking overhead |
| **Voltage Down-Conversion Efficiency** | Source- and topology-dependent | Dedicated charge / voltage conversion for high-voltage TENGs [124,127,128] | Improved high-voltage / high-impedance compatibility |
| **Integration Level** | Separate source-specific conversion stages | Integrated / shared-conversion architectures [97,122,126] | Reduced circuit duplication |

From a system perspective, the relevant optimization target is therefore the net energy delivered to storage and the load over the complete operating cycle, rather than the peak efficiency, minimum startup voltage, or MPPT accuracy of an isolated circuit block.

## 8. Applications and Case Studies: From Laboratory to Field Deployment

The practical significance of micro-energy harvesting is ultimately determined by whether an application-relevant resource can be converted, conditioned, stored and delivered to a functional load under realistic operating constraints. The materials, transducers, hybrid architectures, power-management circuits and storage technologies discussed above therefore converge at the application level, where resource availability, conversion efficiency, form factor, system losses and duty cycle must be considered together. Here, three representative case studies (cardiac monitoring, in-pipe sensing and smart glasses) are used to examine the transition from functional demonstrations toward energy-neutral and long-duration autonomous operation. Importantly, a "self-powered" demonstration does not necessarily imply full energy autonomy, which additionally requires sustained source-to-load balance, storage-state control and minimal external assistance.

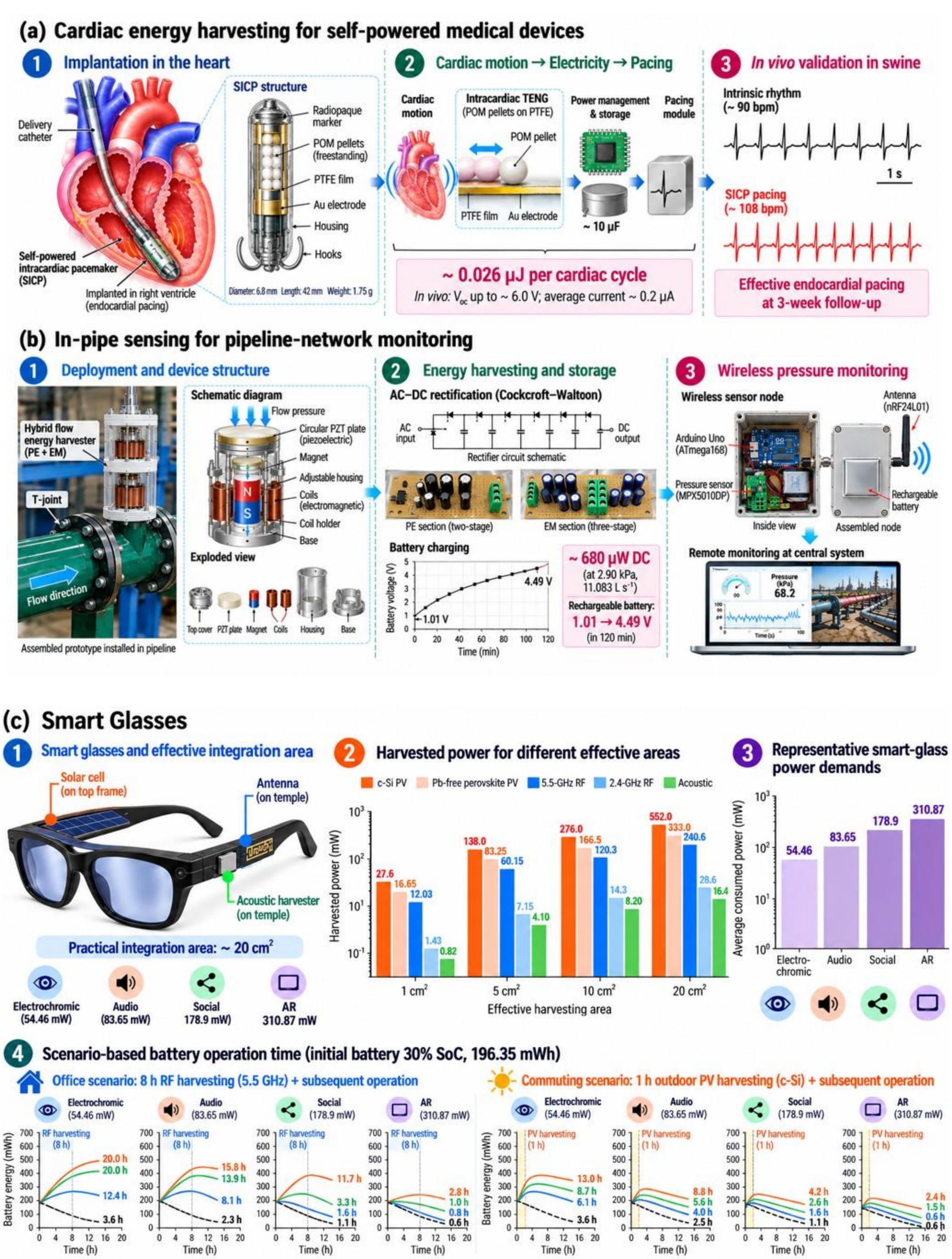
(a) Cardiac energy harvesting for self-powered medical devices
1 Implantation in the heart
Delivery catheter
Self-powered intracardiac pacemaker (SICP)
Implanted in right ventricle (endocardial pacing)
SICP structure
Radiopaque marker
POM pellets (freestanding)
PTFE film
Au electrode
Housing
Hooks
Diameter: 6.8 mm Length: 42 mm Weight: 1.75 g
2 Cardiac motion → Electricity → Pacing
Cardiac motion
Intracardiac TENG (POM pellets on PTFE)
POM pellet
PTFE film
Au electrode
Power management & storage
~ 10 μF
Pacing module
~ 0.026 μJ per cardiac cycle
In vivo: V_oc up to ~ 6.0 V; average current ~ 0.2 μA
3 In vivo validation in swine
Intrinsic rhythm (~ 90 bpm)
1 s
SICP pacing (~ 108 bpm)
Effective endocardial pacing at 3-week follow-up
(b) In-pipe sensing for pipeline-network monitoring
1 Deployment and device structure
Hybrid flow energy harvester (PE + EM)
T-joint
Flow direction
Assembled prototype installed in pipeline
Schematic diagram
Flow pressure
Circular PZT plate (piezoelectric)
Magnet
Adjustable housing
Coils (electromagnetic)
Coil holder
Base
Exploded view
Top cover
PZT plate
Magnet
Coils
Housing
Base
2 Energy harvesting and storage
AC–DC rectification (Cockcroft–Waltoon)
AC input
DC output
Rectifier circuit schematic
PE section (two-stage)
EM section (three-stage)
Battery charging
Battery voltage (V)
Time (min)
1.01 V
4.49 V
~ 680 μW DC
(at 2.90 kPa, 11.083 L s⁻¹)
Rechargeable battery:
1.01 → 4.49 V
(in 120 min)
3 Wireless pressure monitoring
Wireless sensor node
Antenna (nRF24L01)
Arduino Uno (ATmega168)
Pressure sensor (MPX5010DP)
Rechargeable battery
Inside view
Assembled node
Remote monitoring at central system
Pressure (kPa)
68.2
(c) Smart Glasses
1 Smart glasses and effective integration area
Solar cell (on top frame)
Antenna (on temple)
Acoustic harvester (on temple)
Practical integration area: ~ 20 cm²
Electrochromic (54.46 mW)
Audio (83.65 mW)
Social 178.9 mW)
AR 310.87 mW
2 Harvested power for different effective areas
c-Si PV
Pb-free perovskite PV
5.5-GHz RF
2.4-GHz RF
Acoustic
Harvested power (mW)
27.6
16.65
12.03
1.43
0.82
138.0
83.25
60.15
7.15
4.10
276.0
166.5
120.3
14.3
8.20
552.0
333.0
240.6
28.6
16.4
1 cm²
5 cm²
10 cm²
20 cm²
Effective harvesting area
3 Representative smart-glass power demands
Average consumed power (mW)
54.46
83.65
178.9
310.87
Electro-chromic
Audio
Social
AR
4 Scenario-based battery operation time (initial battery 30% SoC, 196.35 mWh)
Office scenario: 8 h RF harvesting (5.5 GHz) + subsequent operation
Electrochromic (54.46 mW)
Audio (83.65 mW)
Social (178.9 mW)
AR (310.87 mW)
RF harvesting (8 h)
Battery energy (mWh)
Time (h)
20.0 h
20.0 h
12.4 h
3.6 h
15.8 h
13.9 h
8.1 h
2.3 h
11.7 h
3.3 h
1.6 h
1.1 h
2.8 h
1.0 h
0.8 h
0.6 h
Commuting scenario: 1 h outdoor PV harvesting (c-Si) + subsequent operation
PV harvesting (1 h)
13.0 h
8.7 h
6.1 h
3.6 h
8.8 h
5.6 h
4.0 h
2.5 h
4.2 h
2.6 h
1.6 h
1.1 h
2.4 h
1.5 h
0.6 h
0.6 h
20 cm²
10 cm²
5 cm²
No harvester

**Figure 15. Case-study deployments: from energy harvesting to functional energy closure.** (a) Cardiac energy harvesting for self-powered medical devices. A battery-free intracardiac triboelectric pacemaker harvested cardiac motion (~0.026 μJ per cardiac cycle) through an integrated energy-harvesting, power-management and storage architecture, enabling effective endocardial pacing in swine at a three-week follow-up [129]. (b) In-pipe sensing: a hybrid piezoelectric–electromagnetic flow harvester delivered up to ~680 μW conditioned DC, charging a 4.8 V storage unit from 1.01 to 4.49 V within 120 min and supporting wireless pressure monitoring [131]. (c) Smart glasses: photovoltaic, RF and acoustic harvesting were evaluated within an approximately 20 $cm^2$ practical integration area against representative application loads, using benchmarks spanning 0.82–27.6 $mW/cm^2$ [132–140]. Scenario-based energy budgeting demonstrates that functional energy closure is governed by the coupled source availability–integration area–storage–load balance, with low-power functions benefiting strongly from larger harvesting areas while continuous high-power AR remains challenging.

### 8.1. Case Study A: Cardiac Energy Harvesting for Self-Powered Medical Devices

**The Environment:** Cardiac energy harvesting presents a stringent test of self-powered medical electronics because the available biomechanical energy is limited and low-frequency, whereas reliable operation must be achieved under severe constraints on device volume, biocompatibility, mechanical stability, and implantation geometry. The relevant benchmark is therefore not peak transducer output, but whether physiological motion can sustain a complete harvesting–power-management–storage–load pathway.

**The Harvesting Solution:** Liu et al. recently demonstrated a battery-free intracardiac pacemaker in swine [129]. The catheter-deliverable capsule (1.75 g, 1.52 $cm^3$) integrates a POM/PTFE triboelectric harvester with rectification, capacitive storage, and pacing electronics. Cardiac motion is converted into electrical energy and accumulated before stimulation, thereby temporally decoupling the weak biomechanical source from the pulsed therapeutic load. Figure 15(a) illustrates the corresponding intracardiac harvesting–storage–pacing architecture.

**Performance:** In vivo, the harvester produced approximately 6.0 V and 0.2 μA, equivalent to 0.026 μJ per cardiac cycle and a maximum output power of 0.039 μW. Despite this sub-μW power level, a 10 μF capacitor charged to 3 V sustained pacing for nearly 40 s, and an atrioventricular-block swine model was successfully paced at approximately 1.5 V. Functionality was maintained over a three-week follow-up, with pacing at week 3 increasing heart rate from approximately 90 to 108 bpm [129]. More recently, Ouyang et al. reported on a symbiotic transcatheter pacemaker employing electromagnetic induction and a magnetic-levitation energy-cache architecture to reduce collision and friction losses and enable a near-zero boot threshold. In a porcine brady-arrhythmia model, regenerated energy exceeded the critical requirement for sustained pacing, enabling simultaneous energy regeneration and therapeutic operation for one month [130].

**Key Takeaway:** These studies shift the design criterion from maximizing instantaneous electrical output to achieving functional energy closure. The Liu's system demonstrates that even sub-μW harvesting can support a pulsed therapeutic load when energy is accumulated and released according to load demand, whereas Ouyang's system advances this principle toward sustained autonomous operation. For implantable cardiac systems, startup threshold, PMIC and storage losses, source–load temporal matching, biocompatibility, and long-term energy balance are therefore as critical as transducer efficiency or peak power.

## 8.2. Case Study B: In-Pipe Sensor Networks for Water Grids

**The Environment:** Pipeline-monitoring networks offer a comparatively persistent source of fluid-flow energy, making local harvesting attractive for sensor nodes where battery replacement is costly or impractical. Unlike implantable systems, however, the central constraint is not only energy scarcity: extracting power from the flow can itself perturb the host infrastructure. Practical viability therefore requires sufficient harvesting–conditioning–storage–sensing–communication closure while minimizing hydraulic and maintenance penalties.

**The Harvesting Solution:** Rahman et al. recently demonstrated an integrated pipeline-monitoring system based on a hybrid piezoelectric–electromagnetic flow energy harvester (HFEH) [131]. A unimorph piezoelectric plate, permanent magnet, and dual electromagnetic coils exploit the same flow-induced excitation through complementary transduction mechanisms, providing a compact route to convert pipeline flow into usable electrical energy. The corresponding hybrid flow-harvesting and wireless-monitoring pathway is presented in Fig. 15(b).

**Performance:** Under controlled conditions of approximately 2.90 kPa flow pressure and 11.08 L $s^{-1}$ flow rate, the hybrid harvester generated up to 831.7 μW AC and 680 μW DC after power conditioning. More importantly, the conditioned output charged a 4.8 V rechargeable storage unit from 1.01 to 4.49 V within 120 min and supported a low-power wireless pressure-monitoring system [131]. The demonstration therefore establishes much of the functional pathway from pipeline flow → hybrid transduction → power conditioning/storage → sensing → wireless communication, rather than reporting transducer output alone.

**Key Takeaway & Caution:** For in-pipe harvesting, the appropriate figure of merit is not maximum electrical output alone but the net energetic and operational benefit to the host infrastructure. Harvested power must therefore be evaluated against pressure loss, flow disturbance, fouling, structural reliability, and maintenance requirements.

## 8.3. Case Study C: Smart Glasses

**The Environment:** Smart glasses provide a stringent testbed for wearable energy harvesting because source availability, integration area, storage capacity and application load are simultaneously constrained. Ambient light, RF radiation and sound offer complementary energy sources, but their availability varies with user activity, whereas the eyewear form factor restricts the usable harvesting area. As illustrated in Fig. 15 (c1), PV elements can be integrated along the frame, with RF antennas and acoustic harvesters accommodated within the temples. Here, an effective harvesting area of up to approximately 20 $cm^2$ was considered practical. The reference 170 mAh, 3.85 V battery corresponds to ~654.5 mWh, with 30% SoC (~196.35 mWh) used as the initial energy for scenario-based analysis.

**The Harvesting Solution:** PV, RF and acoustic harvesting were evaluated using literature-derived performance benchmarks and source-specific modelling (Fig. 15 (c2)). A triboelectric–piezoelectric nanomesh acoustic harvester provides a benchmark of 0.82 mW/$cm^2$ [134], while rectenna and metasurface architectures represent RF harvesting at 2.4 and 5.5/5.8 GHz [135,136], corresponding to modeled benchmarks of 1.43 and 12.03 mW/$cm^2$, respectively. For PV, a Pb-free 2D/3D Sn-halide perovskite cell with a certified PCE of 16.65% provides a benchmark of 16.65 mW/cm² [137], whereas the crystalline-Si research-cell benchmark of 27.6% corresponds to 27.6 mW/cm² under AM1.5G illumination [138]. Assuming ideal area scaling, a 20 cm² harvesting area therefore yields 552, 333, 240.6, 28.6 and 16.4 mW for crystalline-Si PV, Pb-free perovskite PV, 5.5-GHz RF, 2.4-GHz RF and acoustic harvesting, respectively. These values span the representative smart-glass loads from 54.46 mW (electrochromic) to 310.87 mW (AR) (Fig. 15 (c3)), highlighting the strong dependence of energy closure on both source and load.

**Performance:** Scenario-based battery analysis translates this source–area–load balance into operating time (Fig. 15c (4)). In an 8 h office scenario starting from 30% SoC, modeled 5.5-GHz RF harvesting provides 60.15–240.6 mW over 5–20 cm². At 5 cm², it can sustain the electrochromic load and extend audio operation to approximately 8.1 h; at 20 cm², social-glass operation reaches approximately 11.7 h, whereas AR remains limited to ~2.8 h. For a 1 h outdoor commuting scenario, crystalline-Si PV provides 138–552 mW over the same area range. A 10 cm² PV area extends electrochromic operation to ~8.7 h, and 20 cm² extends audio operation to ~8.8 h, whereas social and AR operation reach only ~4.2 and 2.4 h, respectively. Thus, high instantaneous harvested power alone does not guarantee autonomy when source exposure is intermittent and load demand is high.

**Key Takeaway & Caution:** Smart-glass autonomy is governed by the coupled source availability–integration area–storage–load balance, rather than peak harvester output alone. Larger harvesting areas can bring low- and moderate-power functions toward functional energy closure, whereas continuously operating augmented reality (AR) remains difficult within realistic eyewear dimensions. Consequently, hybrid harvesting, duty cycling, low-power electronics and storage-aware power management are likely to be as important as further improvements in transducer power density. These results should nevertheless be interpreted as scenario-based energy-budget projections, not demonstrations of fully autonomous smart glasses. The calculations assume ideal linear area scaling and neglect PMIC, conversion, storage and interconnection losses; real-world performance will additionally depend on source intermittency, illumination angle and shading, device placement, optical and mechanical constraints, and user comfort.

A cross-case comparison of the three application scenarios is provided in Table 7. Despite the large differences in harvested power, source availability, form factor, and load demand, the case studies converge on a common system-level requirement: practical autonomy is determined by the complete balance among ambient resource availability, transducer output, power-management losses, storage capacity, load duty cycle, and operating duration, rather than by peak harvester performance alone.

**Table 7. Case Study Comparison: Deployment Performance and Lessons Learned**

| Parameter | Case A: Cardiac Implant | Case B: In-Pipe Sensor | Case C: Smart Glasses |
|---|---|---|---|
| **Application** | Cardiac pacing [129,130] | Water-grid pressure monitoring [131] | Smart-glass energy autonomy [132–140] |
| **Harvester Type** | Triboelectric / electromagnetic [129,130] | Piezoelectric–electromagnetic [131] | PV / RF / acoustic [134–138] |
| **Harvester Footprint** | 1.52 cm³; 1.75 g [129] | Pipe-mounted [131] | Up to ~20 cm² considered* |
| **Primary Energy Sources** | Cardiac motion | Pipeline fluid flow | Ambient light / RF / sound |

| **Average Harvested Power** | 0.026 µJ/cycle; max. 0.039 µW [129] | 680 µW conditioned DC [131] | Scenario-dependent; up to 240.6 mW RF or 552 mW PV at 20 $cm^2$* |
|---|---|---|---|
| **Peak Harvested Power** | ~6.0 V, 0.2 µA in vivo [129] | 831.7 µW AC [131] | Component benchmarks: 0.82–27.6 $mW/cm^2$ [134–138] * |
| **Load** | Pacing electronics | Wireless pressure monitoring | Electrochromic / audio / social-glass / AR |
| **Transmission Interval** | Energy accumulated before pacing | After storage charging | Load-dependent* |
| **Deployment Duration** | 3 weeks [129]; 1 month [130] | Controlled demonstration [131] | Scenario-based projection* |
| **Primary Failure Mode** | Biocompatibility / long-term reliability | Fouling / hydraulic effects | Source intermittency / limited area |
| **Critical Engineering Compromise** | Energy supply vs. implant constraints | Harvested power vs. hydraulic penalty | Harvesting area vs. wearability |
| **Key Insight** | Energy buffering enables pulsed therapy | Net benefit must include hydraulic penalties | Peak power ≠ energy autonomy |
| **Lifetime-Limiting Component** | Implant / storage / interface | Mechanical components / storage | Storage / PMIC |
| **Cost-to-Power Ratio** | High | Infrastructure-dependent | Application-dependent |
| **Technology Readiness Level (TRL)** | Preclinical in-vivo demonstration | Integrated controlled demonstration | Scenario / component-level analysis* |
| **Next Step** | Long-term reliability / clinical translation | Field reliability / hydraulic assessment | System integration / energy closure |

Note: * *Smart-glass values are scenario-based estimates derived from literature benchmarks and ideal area scaling rather than measurements from an integrated autonomous smart-glass system.*

The comparison also highlights the need to distinguish between energy harvesting, functional self-powered operation, and sustained energy autonomy. The cardiac and in-pipe systems experimentally demonstrate substantial portions of the source–harvester–power-management–storage–load pathway, whereas the smart-glass case primarily establishes a scenario-based energy budget under defined source, area, storage, and load assumptions. Thus, a device should not be considered fully autonomous solely because its harvester can momentarily exceed the load power. Long-term autonomy instead requires a positive energy balance over the relevant operating cycle while accounting for source intermittency, conversion losses, storage leakage, and application-specific duty cycles.

## 9. Final Synthesis and Future Outlook

Across mechanical, thermal, radiant, and hybrid energy-harvesting technologies, a common transition is emerging from isolated transducer optimization toward complete self-powered microsystems. Practical autonomy can no longer be judged by peak voltage, current, or power density alone; rather, it depends on the full energy pathway, from environmental resource capture and transduction to power management, storage, regulation, and the time-varying demands of sensing, computation, communication, or actuation. Recent implantable systems further demonstrate that energy conversion, storage, miniaturization, biocompatibility, and system integration must be addressed as a coupled design problem [139]. Therefore, four overarching conclusions define this transition from laboratory harvesters toward practically deployable autonomous systems. Continued materials innovation remains important, but substantial advances have already been achieved through charge engineering, nonlinear mechanics, bandgap optimization, interfacial regulation, and hybridization. Increasingly, practical improvement depends on how efficiently harvested energy is conditioned, stored, and converted into useful service. Recent self-powered implants illustrate this transition from isolated transducers toward architectures integrating energy regeneration, power conditioning, storage, and therapeutic function [130,139]. Future progress will therefore depend increasingly on system-level co-design across the entire energy pathway. A high-performance harvester does not ensure autonomous operation if harvested energy cannot be efficiently accumulated and delivered at the voltage and timescale required by the load. This mismatch is particularly critical when weak, continuous harvesting must sustain intermittent burst loads such as system startup or wireless transmission [27]. Storage capacity, ESR, leakage, usable voltage window, cycling stability, recharge time, and PMIC quiescent consumption can therefore become more restrictive than transducer output. Meaningful cross-study comparison remains difficult because environmental inputs and electrical measurement boundaries are often non-equivalent. Mechanical harvesters require defined excitation conditions; thermal devices require spatial or temporal temperature profiles; indoor photovoltaics require spectrum-aware irradiance; and Ambient RF and Backscatter systems require defined electromagnetic and receiver conditions. Likewise, open-circuit voltage, matched-load power, rectified output, stored energy, and regulated load power represent different stages of the energy pathway. Standardization should therefore define both the applied resource and the measurement boundary through modality-specific protocols. Existing guidance for indoor photovoltaics [106] and recent characterization efforts for piezoelectric biomaterials [140] provide useful precedents. The primary value of micro-energy harvesting is not replacing conventional power generation, but reducing battery replacement and external maintenance in inaccessible, distributed, wearable, structural, and implantable systems. This shifts the optimization target from maximum instantaneous power toward retained useful service over application-relevant lifetimes. Ultimately, the most valuable harvester is one that enables reliable system function with minimal external intervention. Taken together, the advances reviewed here indicate a broader maturation of energy harvesting beyond proof-of-concept transducer demonstrations toward system-level autonomy. The central question is therefore shifting from how much power can a harvester generate? to how much useful service can the complete system sustainably deliver?

First, transducer-level performance has advanced substantially, but does not by itself establish practical autonomy. Charge-pumped TENGs increase attainable surface charge density [39,40], while nonlinear and bistable piezoelectric harvesters broaden operating bandwidth [64,66]. Cross-technology ranking nevertheless remains intrinsically limited because mechanical, thermal, optical, and RF harvesters operate from physically distinct resources, while open-circuit, matched-load, rectified, stored, and regulated outputs represent different stages of the energy pathway.

Second, increasingly important gains arise from architectural innovation rather than incremental material optimization alone. Liquid–solid and liquid-metal triboelectric systems reduce dependence on abrasive solid–solid contact [43–45], nonlinear architectures extend operation beyond narrow resonance [64,66], and hybrid systems exploit complementary resources, shared structures, or functional coupling [86–88,91,94]. Such approaches change how environmental energy is captured, conditioned, and used, providing a bridge from transducer optimization toward system integration.

Third, hybridization should be judged by demonstrated functional and system-level benefit rather than source count. Temporal complementarity, shared structures, and functional coupling are advantageous only when their benefits exceed additional PMIC, control, leakage, and reliability costs. FusedAR exemplifies such co-design by using solar and kinetic harvesters simultaneously for energy acquisition and activity sensing, enabling energy-positive human-activity recognition [144]. Taken together, these findings establish a common design principle: optimize the stage that currently limits useful system service rather than maximizing transducer performance in isolation.

### 9.1. Critical Gaps and Unresolved Challenges

Despite substantial progress, several unresolved system-level gaps continue to limit the translation of high-performance harvesters into autonomous microsystems. Three challenges are particularly consequential.

Gap 1: The Absence of Standardized Testing Protocols. The lack of harmonized testing remains a fundamental barrier to meaningful cross-study comparison. Mechanical harvesters tested under different excitation spectra, forces, or environmental conditions cannot be directly compared, just as indoor photovoltaic devices characterized under different illumination spectra cannot be ranked using illuminance alone. Standardization must therefore define both the environmental input and the electrical measurement boundary. Mechanical harvesters require defined excitation and boundary conditions; thermoelectric and pyroelectric devices require spatial or temporal temperature profiles; indoor photovoltaics require spectral irradiance; and RF harvesters require defined frequency, polarization, field conditions, antenna coupling, and accepted power. Likewise, open-circuit, matched-load, rectified, stored, and regulated-load outputs should be explicitly distinguished. Existing indoor-photovoltaic guidance [106] and reproducible characterization approaches for piezoelectric biomaterials [140] provide useful precedents. Extending such modality-specific protocols across energy-harvesting technologies is essential for credible benchmarking, reproducibility, and eventual system-level comparison.

Gap 2: Long-Term Reliability Data is Practically Non-Existent. Long-duration reliability remains substantially less established than short-term transducer performance. Mechanical fatigue, triboelectric-interface evolution, liquid-metal oxidation, thermal cycling, encapsulation degradation, storage aging, and environmental exposure can progressively alter the energy pathway and shift the dominant system bottleneck. Recent implantable studies represent important progress. A symbiotic transcatheter pacemaker demonstrated one month of autonomous operation in a porcine disease model [130], while a TENG-powered leadless intracardiac pacemaker demonstrated stable in-vivo energy harvesting and pacing [141]. These results strengthen the evidence for integrated self-powered operation but do not yet establish multi-year clinical reliability. Future studies should therefore move beyond isolated cycle counts and short-term output retention toward accelerated aging, environmental exposure, encapsulation and interface degradation, storage aging, and failure-mode analysis. Ultimately, reliability should be quantified as retained useful system service under application-relevant conditions, rather than retained transducer output alone.

Gap 3: The Impedance Matching and Heat Dissipation Dilemma for Hybrids. Hybrid and multi-source systems introduce a fundamental interface challenge because heterogeneous harvesters can differ by orders of magnitude in voltage, current, source impedance, intermittency, and optimal extraction strategy. TENGs, thermoelectric generators, photovoltaic devices, piezoelectric harvesters, and RF rectennas therefore require source-specific extraction, impedance conversion, isolation, and control. The PMIC approaches discussed in Section 7, including sub-100-mV cold start, low-overhead MPPT, ultra-low quiescent current, and synchronized multi-input extraction, provide key enabling functions, but combining additional sources does not inherently improve system performance. Hybridization is beneficial only when the additional usable energy exceeds the switching, sensing, control, leakage, and conversion losses introduced by the interface. Thermal management creates a parallel constraint in compact wearable and implantable systems, where losses from power conversion, regulation, computation, and communication are concentrated within small thermal volumes [145]. Effective multi-source integration therefore requires coordinated co-design of heterogeneous harvesters, power electronics, storage, workload, and thermal pathways.

### 9.2. The 2030 Roadmap: A Projection

The available evidence supports a roadmap toward 2030 centered on energy-system closure rather than peak transducer output, but not deterministic commercialization timelines (Fig. 16). The most recent literature reviewed here indicates a transition toward hybrid integration, energy storage, low-voltage power management, and standardized system-level evaluation. Building on this foundation, three research directions are particularly relevant toward 2030.

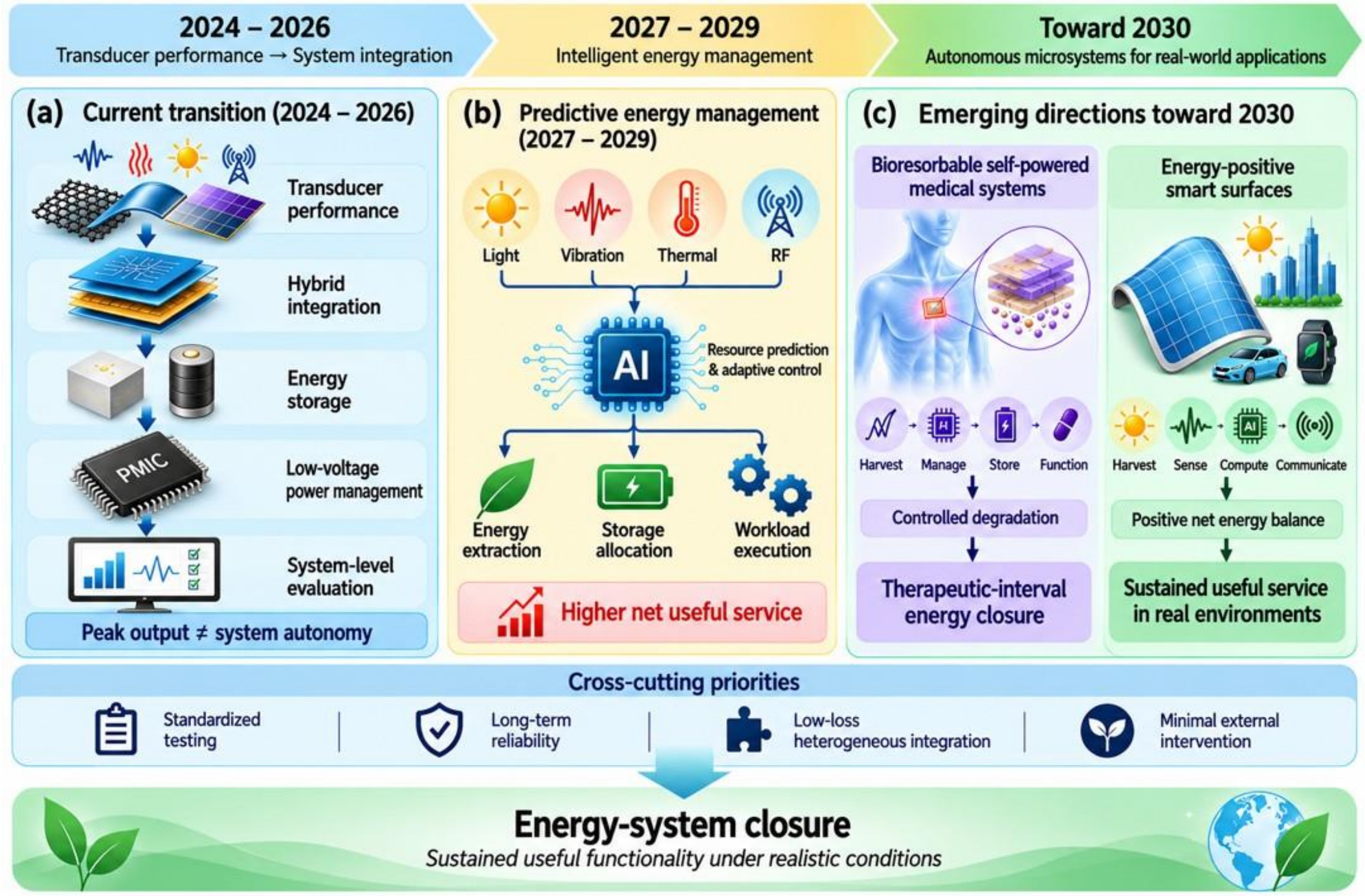

**Figure 16. 2030 technology roadmap: from peak output to energy-system closure.** (a) Current transition (2024–2026): shifting from peak transducer performance toward hybrid integration, energy storage, low-voltage power management, and system-level evaluation. (b) Predictive energy management: resource-aware scheduling and ultra-low-overhead control coordinate energy extraction, storage, and workload execution [143]. (c) Emerging directions toward 2030: bioresorbable self-powered medical systems

integrating harvesting, storage, power management, and functional devices [72,139,140,146], and energy-positive smart surfaces combining harvesting, sensing, computation, and communication [144,147,148]. Cross-cutting priorities include standardized testing, long-term reliability, low-loss heterogeneous-source integration, and sustained useful operation with minimal external intervention, collectively advancing energy-system closure.

AI-Driven Autonomous Power Management: Energy-aware scheduling already demonstrates that workload can be adapted to time-varying environmental energy [143]. A logical next step is predictive resource-aware management, in which light, vibration, RF, thermal conditions, or storage trajectories are used to anticipate resource-rich and resource-poor periods and dynamically adjust extraction, storage allocation, and task execution. Machine-learning-assisted design and self-powered sensing discussed earlier [49–51] provide enabling concepts, but intelligence carries its own sensing, memory, computation, and switching costs. An AI-enabled PMIC should therefore be evaluated by whether predictive management increases net useful service after all control overhead is included. The relevant goal toward 2030 is thus not simply to embed AI in the PMIC, but to develop ultra-low-overhead predictive management that demonstrably improves energy neutrality, service availability, or resilience to resource drought.

Truly Biodegradable and Bioresorbable Harvesters for the Medical Market: Biodegradable piezoelectric materials, including PLLA and naturally derived structures, provide promising transduction pathways for temporary implants [72,140], while recent studies emphasize the need to co-optimize miniaturization, biocompatibility, conversion efficiency, and system integration [139]. A truly bioresorbable self-powered system, however, requires compatible interconnects, rectifiers, PMICs, storage, functional electronics, and encapsulation—not merely a degradable harvester. Although preclinical bioresorbable platforms demonstrate progress [146], complete self-powered operation remains to be established. The appropriate roadmap objective is therefore bioresorbable energy-system closure: sustained function over the required therapeutic interval followed by controlled, biologically acceptable degradation.

The "Energy-Positive" Smart Surface: Flexible and conformal surfaces integrating environmental harvesting, sensing, computation, and communication represent another route toward autonomous operation across wearables, vehicles, buildings, and distributed infrastructure. Such systems should be considered energy-positive only when they complete their declared sensing, inference, communication, or control workload while maintaining a positive net energy balance after conversion, storage, sensing, computation, and communication losses are included. FusedAR provides an early system-level example by using solar and kinetic harvesting signals for both energy acquisition and activity sensing, enabling energy-positive human-activity recognition [144]. The subsequent SKEH-HAR dataset extends this solar–kinetic energy/sensing concept toward reproducible multimodal evaluation [147]. More broadly, intelligent self-powered sensing and AIoT architectures [148] may reduce communication demand through local inference or support adaptive operation, provided that computational overhead is included in the energy balance. Across these directions, standardized testing, long-term reliability, low-loss heterogeneous-source integration, and sustained useful operation with minimal external intervention remain cross-cutting priorities. The central design principle toward 2030 is therefore to optimize the stage that constrains useful service under the declared operating condition, treating materials, harvesters, PMICs, storage, sensing, computation, and communication as coupled elements of a single energy pathway rather than ranking them in isolation. Progress toward reliable autonomous microsystems will ultimately be determined by energy-system closure—the sustained delivery of useful functionality under realistic environmental conditions with minimal external intervention over the required lifetime—rather than by peak harvested power alone.

## 10. Conclusion

The field of energy harvesting has undergone a remarkable evolution in the most recent period, transitioning from a disparate collection of physics curiosities into a coherent, multidisciplinary engineering discipline that intersects materials science, semiconductor physics, power electronics, and data science. The foundational principles are now robustly understood; the materials are increasingly scalable; and the power management ICs are steadily closing the efficiency gap. However, the ultimate triumph of energy harvesting, the realization of a trillion-node, battery-free IoT ecosystem, depends not on a single breakthrough, but on the system-level orchestration of all the domains reviewed herein. It demands that the materials scientist, the circuit designer, and the algorithm developer work in lockstep, guided by real-world deployment data rather than idealized laboratory metrics. We are optimistic that the next five years will witness the maturation of this ecosystem, ushering in an era of truly autonomous, maintenance-free, and environmentally sustainable microsystems that will fundamentally reshape how we interact with the physical and digital worlds.